\documentclass[14pt,epsfig]{article}
\usepackage{amsfonts}
\usepackage{color}
\usepackage{amsmath}
\usepackage{amssymb}
\usepackage{graphicx}
\usepackage{float}
\usepackage{titlesec}

\titleformat*{\section}{\large\bfseries}
\titleformat*{\subsection}{\bfseries}

\newtheorem{prop}{Proposition}[section]

\newtheorem{theorem}{Theorem}[section]
\newtheorem{remark}[theorem]{Remark}

\newtheorem{lemma}[theorem]{Lemma}
\newtheorem{proposition}[theorem]{Proposition}

 \def\2{I$\!$I}

\usepackage{color}
\newcommand{\1}{1\!\!1}

\def\*{{\phantom *}}
\newcommand{\N}{\Bbb N}

\newcommand{\R}{\Bbb R}

\newcommand{\bml}[1]{\begin{multline}\label{#1}}
\newcommand{\bee}{\begin{equation}}
\newcommand{\bed}{\begin{displaymath}}
\newcommand{\ee}{\end{equation}}

\newcommand{\bs}{\begin{split}}

\newcommand{\be}{\beta}
\newcommand{\ga}{\gamma} \newcommand{\Ga}{\Gamma}
 
\newcommand{\la}{\lambda} \newcommand{\La}{\Lambda}

\newcommand{\si}{\sigma}

\newcommand{\dist}{\operatorname{dist}}

\newcommand{\w}{\widetilde}

\newcommand{\ov}{\overline}
\newcommand{\non}{\nonumber}

\begin{document}

\begin{center}
\renewcommand{\thefootnote}{\fnsymbol{footnote}}
{\Large \textbf{Correlation of clusters: Partially truncated correlation functions and their decay.}}

\vspace{0.5cm}
{\textbf{T.~C.~Dorlas$^1$, A.~L.~Rebenko$^2$, B.~Savoie$^1$}}
\date{}
\begin{footnotesize}
\begin{tabbing}
$^1$ \= Dublin Institute for Advanced Studies, School of Theoretical Physics, 10 Burlington Road, Dublin 04, Ireland.\\
$^2$ \> Institute of Mathematics, Ukrainian National Academy of Sciences, Kyiv, Ukraine.
\end{tabbing}
\end{footnotesize}


\setcounter{footnote}{0} \renewcommand{\thefootnote}{\arabic{footnote}}


\vspace{0.5cm}
\textbf{Abstract} \\[0pt]
\end{center}
In this article, we investigate partially truncated correlation functions (PTCF) of infinite continuous systems of classical point particles with pair interaction. We derive Kirkwood-Salsburg-type equations for the PTCF and write the solutions of these equations as a sum of contributions labelled by certain \textit{forest graphs}, the connected components of which are tree graphs. We generalize the method developed by R.A. Minlos and S.K. Poghosyan (1977) in the case of truncated correlations. These solutions make it possible to derive strong cluster properties for PTCF which were obtained earlier for lattice spin systems.
\vspace{0.5cm}

\noindent \textbf{Keywords:} Classical statistical mechanics,
 strong cluster properties, truncated  correlation functions.
\\
\noindent \textbf{Mathematics Subject Classification 2010:} 82B05, 82B21.

\medskip

\textheight 24.0cm \textwidth 16.0cm \headheight 0cm \headsep 0cm \footskip %
1cm \topmargin 0cm


\tableofcontents
\medskip

\section{Introduction.}

\setcounter{equation}{0} \renewcommand{\theequation}{\arabic{section}.%
\arabic{equation}} 

Correlation functions were first introduced in Statistical Mechanics by L.S. Ornstein and F. Zernike at the beginning of the 20th century in the study of critical fluctuations, see \cite{OZ26}. Mathematical studies apparently began with the work of J. Yvon \cite{Yv35} and the independent works of N.N. Bogolyubov \cite{Bog46}, J.G. Kirkwood \cite{Ki46}, and,  M. Born and H.S. Green \cite{BG47}. In some sense, they were completed in the works of O. Penrose \cite{Pe63}, D. Ruelle \cite{Ru69}, and, N.N. Bogolyubov \textit{et al.} \cite{BPK69}. Correlation functions are the densities of correlation measures and were called m-particle distribution functions by N.N. Bogolyubov, which more accurately describes their meaning.  The physical correlations between particles are in fact described by the so-called \textit{truncated correlation functions} (TCF), or connected correlation functions, which become zero in the absence of interaction between the particles.\\
\indent When studying the thermodynamic properties of statistical systems, the important characteristics are often interactions between groups of particles (the so-called clusters). Correlations between clusters are described by the so-called \textit{partially truncated correlation functions} (PTCF), or partially connected correlation functions. In \cite{Le72}, J.L. Lebowitz derived bounds on the decay of correlations between
two widely separated sets of particles (two point-PTCF) for ferromagnetic Ising spin systems in terms of the decay of the pair correlation. Later, in \cite{DIS73}, some 'physically reasonable' hypotheses on the decay of the TCF and PTCF were presented and discussed. In subsequent publications of these authors \cite{DIS74, DS76}, various strong decay properties were proved for TCF of lattice and continuous systems in different situations. In \cite{IS78}, some general results on strong cluster properties of TCF and PTCF for lattice gases are presented
(in fact, the proof of their main theorem involves long technical parts which were obtained in unpublished
work of one of the authors).\\
\indent In this paper, we consider classical  continuous  systems of point particles which interact through
a two-body interaction potential. We derive equations of Kirkwood-Salsburg-type for the PTCF and apply
the technique that was proposed by R.A. Minlos and S.K. Poghosyan in \cite{MP77} to obtain solutions of these equations in the form of a series of contributions of certain forest diagrams. Such a representation makes it possible to obtain strong cluster properties for the PTCF in a convenient form for deriving estimates. We stress the point that explicit formulas for the upper bounds are obtained, some of which rely on some original (to our best knowledge) combinatorial identities.


\section{Mathematical background.}

\setcounter{equation}{0} \renewcommand{\theequation}{\arabic{section}.%
\arabic{equation}} 

\subsection{Configuration spaces.}

Let ${\R}^{d}$ be a $d$-dimensional Euclidean space, $d \geq 1$. By $\mathcal{B}({\R}^{d})$ we denote the family of all Borel sets in $\mathbb{R}^{d}$ and by $\mathcal{B}_{c}({\R}^{d})$ the system of all sets in $\mathcal{B}({\R}^{d})$ which are bounded.\\
\indent The  positions $\{x_i\}_{i\in\N} $ of identical particles are assumed to form  a locally finite subset in ${\R}^d$. Because the particles are assumed to be identical, the ordering is irrelevant. Moreover, there can be more than one particle at any point. The configuration space is therefore given by locally finite maps
\begin{equation*}
\Ga=\Gamma_{\R^{d}} := \{\gamma: {\R}^{d}\to \N_0\,:\, \sum_{x \in \Lambda} \gamma(x) <\infty\,\,\textrm{for all}\,\,\Lambda \in \mathcal{B}_c({\R}^{d})\},
\end{equation*}
where we set $\N_0 := \N \cup \{0\}$. For any $\La\in\mathcal{B}_c(\R^d)$, we hereafter denote by $\ga_\La$ the restriction of $\ga$ to $\La$. Further, we define the space of finite configurations $\Ga_{0}$ in $\R^d$ as
\begin{equation*}
\Ga_{0} := \bigsqcup_{n\in \N_0} \Ga^{(n)},\quad \Ga^{(n)}:= \{\gamma \in\Ga\, :\, \sum_{x \in \R^d} \gamma(x) =n \},
\end{equation*}
and the space of finite configurations in $\La$ as
\begin{equation*}
\Gamma_{\La} :=  \bigsqcup_{n\in \N_0} \Ga^{(n)}_\La,\quad \Ga^{(n)}_\La:=\{\gamma\in\Ga\,:\, \sum_{x \in \La} \gamma(x) =n,\, \sum_{x \in \La^c} \gamma(x) = 0 \}.
\end{equation*}
The topology on $\Gamma$ is generated by the subbasis $\{\mathcal{O}_K^m\}$, where $m \in \N_0$ and $K$ runs over compact subsets of $\R^d$ with nonempty interior, given by, see, e.g., \cite[Sec. 5]{Ru70},
\begin{equation*}
\mathcal{O}_K^m := \{\gamma \in \Gamma\,:\,\sum_{x \in K} \gamma(x) = \sum_{x \in {\rm Int}(K)} \gamma(x) = m \}. \end{equation*}
The topological space $\Gamma$ is a polish space (i.e., metrizable, separable and complete). The corresponding Borel $\sigma$-algebra $\mathcal{B}(\Ga)$ is generated by the sets
\begin{equation*}
\mathcal{W}_\La^m := \{\gamma \in \Ga\,:\, \sum_{x \in \La} \ga(x) = m \}, \quad \La \in \mathcal{B}_c({\R}^{d}). \end{equation*}
For further details, we refer the readers to \cite{Min68,Le75} and also the later works \cite{KoKu02,KK06}.

\subsection{Poisson measure on configuration spaces.}

States of an {\it ideal gas} in equilibrium statistical mechanics are described by a {\it Poisson random point measure} $\pi_{z\sigma}$ on the configuration space $\Ga$, where $z>0$ is the activity (determining the density of particles) and $\sigma$ denotes the Lebesgue  measure on $\R^d$, i.e., $\sigma(dx)=dx$. So $\pi_{z\sigma}$ is the Poisson measure with intensity measure $z\sigma$. To define $\pi_{z\sigma}$ on $\Ga$, we first introduce a {\it Lebesgue-Poisson measure} $\la_{z\sigma}=\la_{z\sigma}^\La$ on the space of finite configurations $\Ga_\La,\;\La\in\mathcal{B}_c({\R}^{d})$ or $\Ga_0$, see, e.g., \cite{Min68}.
Given an $n$-tuple $(x_1,\dots,x_n) \in \La^n$, define
\begin{equation}
\label{gammax}
\ga_{(x_1,\dots,x_n)}(x) := \sum_{i=1}^{n} \mathbf{1}_{\{x_{i}\}}(x),
\end{equation}
which is independent of the order of the points $x_1,\dots,x_n$. Given a continuous function $F: \Gamma \to \R$, we can put $F_n(x_1,\dots,x_n) := F(\ga_{(x_1,\dots,x_n)})$, $n\in \mathbb{N}$ which defines a continuous symmetric function $F_n: \R^{nd} \to \R$. Then, we define,
\begin{equation}
\label{IL-P}
\begin{split}
\int_{{\Ga}_\La}F(\ga)\,\la_{z\si}(d\ga):={}&\sum_{n=0}^\infty\frac{z^n}{n!}\int_\La\cdots\int_\La
F(\ga_{(x_1,\dots,x_n)})  \,dx_1 \cdots dx_n \\
={}& \sum_{n=0}^\infty\frac{z^n}{n!}\int_\La\cdots\int_\La F_n(x_1,\dots,x_n)\,dx_1\cdots dx_n,
\end{split}
\end{equation}
where the term $n=0$ in the sum is set to $1$ by convention. It can be seen from \eqref{IL-P} that the family of probability measures
\begin{equation*}
\pi_{z\sigma}^\La:=\mathrm{e}^{-z\si(\La)}\la_{z\sigma}^\La,\quad \La\in
\mathcal{B}_c({\R}^{d}),
\end{equation*}
is consistent (i.e., forms a projective system), and by standard arguments,
one can prove that there exists a unique probability  measure $\pi_{z\sigma}$ on the configuration space $\Ga$ which is the projective limit of $\pi_{z\sigma}^\Lambda$.\\
\indent The main feature of the measures $\pi_{z\sigma}$ and $\la_{z\sigma}$  is the independence of restrictions to disjoint Borel sets, which is called infinite divisibility, see, e.g., \cite[Sec. 4.4]{GV68}.
This means that, for example, in the configuration space $\Ga_\La$, the following lemma holds.

\begin{lemma}
\label{infdiv1}
Let $\Lambda \in \mathcal{B}_c(\R^d)$ and $X_k \in\mathcal{B}_c(\R^d)$, $k=1,2$ such that $X_1\cap X_2=\emptyset$ and $X_1\cup X_2= \La$. Then, for all measurable functions $F_k: \Ga_{X_k} \to \R$, the following identity holds
\begin{equation*}
\int_{\Ga_\La}F_1(\ga)F_2(\ga)\,\la_{z\si}(d\ga) = \int_{\Ga_{X_1}}F_1(\ga)\,\la_{z\si}(d\ga)\int_{\Ga_{X_2}}F_2(\ga)\,\la_{z\si}(d\ga).
\end{equation*}
\end{lemma}

In \cite{Re98, PR07, PR09} this property is the main technical tool in proving the existence of correlation functions in the infinite-volume limit. The following identity, which is related to the multivariate Campbell-Mecke formula for point processes, see, e.g., \cite{LP}, is similar and will be used extensively.

\begin{lemma}\label{combresum}
Given $\Lambda \in \mathcal{B}_c(\R^d)$ and for all positive measurable functions $F: \Ga_\La \to \R$ and $H: \Ga_\La\times\Ga_\La \to \R$, the following identity holds
\begin{equation}
\label{equ}
\int_{\Ga_\La} F(\ga) \sum_{\eta\leq\ga} H(\eta,\ga-\eta)\,\la_\si(d\ga)=
\int_{\Ga_\La} \int_{\Ga_\La} F(\eta+\ga) H(\eta,\ga)\,\la_\si(d\eta)\,\la_{\si}(d\ga).
\end{equation}
\end{lemma}

\noindent \textbf{Proof.} Set $d^{n}x:=dx_1\cdots dx_n$. By \eqref{IL-P}, the left-hand side can be rewritten as
\begin{align*}
& \sum_{n=0}^\infty \frac{1}{n!} \int_{\La^n} F_n(x_1,\dots,x_n) \sum_{I \subset \{1,\dots,n\}} H_{|I|,n-|I|}(x_I,x_{I^c})\, d^n x \nonumber \\
 = &\sum_{n=0}^\infty \frac{1}{n!} \sum_{m=0}^n {n \choose m} \int_{\La^n}
 F_n(x_1,\dots,x_m,x_{m+1},\dots,x_n) H_{m,n-m}(x_1,\dots,x_m,x_{m+1},\dots,x_n)\,d^n x \nonumber \\
 = &\sum_{m=0}^\infty \sum_{k=0}^\infty \frac{1}{m!\,k!} \int_{\La^m} \int_{\La^k} F_{m+k} (x_1,\dots,x_m,y_1,\dots,y_k)H_{m,k}(x_1,\dots,x_m,y_1,\dots,y_k)\,d^m x\, d^k y,
\end{align*}
where we set $I^c := \{1,\dots,n\}\setminus I$. It remains to use \eqref{IL-P} again, and \eqref{equ} follows.  $\blacksquare$


\subsection{Distributions in $\mathcal{D'}(\Gamma_0)$.}
\label{rsecto}

The space of test functions $\mathcal{D}(\Gamma_0)$ consists of functions $F:\Gamma_0 \to \R$ given by a sequence $(F_{n})_{n \in \mathbb{N}}$ of symmetric functions $F_n \in C_{0}^\infty(\R^{dn})$ with common support such that
\begin{equation*}
F(\gamma)= F(\ga_{(x_1,\dots,x_n)}) = F_n(x_1,\dots,x_n),\quad \textrm{for any $\gamma\in\Gamma^{(n)}$,}
\end{equation*}
where $\gamma_{(x_{1},\dots,x_{n})}$ is defined as in \eqref{gammax}. Hereafter, we denote $\vert \ga \vert := \sum_{x \in \R^d} \ga(x)$, $\ga \in \Gamma_0$.\\
\indent For a given  $j\in C^\infty_0(\R^d)$ with $\vert j\vert\leq 1$, we introduce the function $\chi_j: \Ga_0\rightarrow\R$ defined as
\begin{align}
\label{LPE}
\eta\mapsto \chi_j(\eta):= \left\{ \begin{array}{ll}
1,  & \eta =\emptyset, \\
\prod\limits_{x\in\eta} j(x),  & |\eta|\geq 1.
\end{array}\right.
\end{align}
Here and hereafter, $x \in \eta$ means $x \in \mathbb{R}^{d}$ such that $\eta(x) \geq 1$. Clearly, $\chi_j \in \mathcal{D}(\Gamma_0)$.\\
\indent For any $\eta\in\Gamma_0$, we define distributions $\delta_\eta$ such that, for any $F\in\mathcal{D}(\Gamma_0)$,
\begin{equation}
\label{DF}
\langle \delta_\eta, F \rangle := \int_{\Gamma_0}\delta_\eta(\gamma)F(\gamma)\,\lambda_{z\sigma}(d\gamma)=z^{|\eta|}F(\eta).
\end{equation}
In terms of 'ordinary' distributions, this means that
\begin{align*}
\delta_\eta(\gamma)
= \left\{ \begin{array}{ll}
0,  & \mbox{if\,\, $\vert\gamma\vert\neq\vert\eta\vert$}, \\
1,  & \mbox{if\,\, $\gamma=\eta =0$}, \\
\sum\limits_{\pi \in \mathcal{S}_m} \prod\limits_{k=1}^m \delta(x_k-y_{\pi(k)}),  & \mbox{if\,\, $\ga =\ga_{(x_1,\dots,x_m)},\,\eta = \ga_{(y_1,\dots,y_m)}$,}
\end{array}\right.
\end{align*}
where $\mathcal{S}_{m}$ is the group of permutations of $\{1,\ldots,m\}$, and the product is a direct product of $\delta$-functions. Note that, if $\eta_1\cdot\eta_2 = 0$ for some $\eta_1, \eta_2 \in \Gamma_0$ then $\delta_{\eta_1}$ and $\delta_{\eta_2}$ commute. In the language of sets this means
that when $\eta_1\cap\eta_2=\emptyset$ the product of distributions $\delta_{\eta_1}\delta_{\eta_2} = 0$ due to \eqref{DF}. So for given collections $(\eta_i)_{i=1}^m$ of $\eta_i \in \Ga_0$ with $\eta_i\cdot\eta_{i'}=0$ if $i\neq i'$ and complex numbers $(\alpha_i)_{i=1}^m$, we can then define the product
\begin{equation}
\label{prd1}
\prod_{i=1}^m\Delta_{(\alpha_i,\eta_i)}(\gamma):=\prod_{i=1}^m(1+\alpha_i \sum_{\xi_i\leq\ga}\delta_{\eta_i}(\xi_i)).
\end{equation}
Note that, if $\eta_i\cdot\eta_{i'} = 0$ for $i\neq i'$, then
\begin{equation*}
\prod_{i\in I} \sum_{\xi_{i} \leq \gamma} \delta_{\eta_{i}}(\xi_{i}) = \sum_{\xi \leq \gamma} \delta_{\sum\limits_{i \in I} \eta_{i}}(\xi),\quad I \subset \{1,\dots,m\}.
\end{equation*}
In distributional form, we have
\begin{equation}
\label{Deltadistr}
\langle \prod_{i=1}^m \Delta_{(\alpha_i,\eta_i)}, F \rangle = \sum_{I \subset \{1,\dots,m\}}
\prod_{i \in I} \alpha_i z^{\vert\eta_i\vert} \int_{\Ga_0} F (\sum_{i \in I} \eta_i + \ga )\,\la_{z\si}(d\ga).
\end{equation}
Indeed, by \eqref{equ} (in distributional form) along with \eqref{DF},
\begin{equation*}
\begin{split}
\int_{\Ga_0}F(\ga)\sum_{\xi \leq \ga} \delta_{\sum\limits_{i \in I} \eta_{i}}(\xi)\,\la_{z\si}(d\ga) &=
\int_{\Ga_0}\int_{\Ga_0}F(\xi+\ga) \delta_{\sum\limits_{i \in I} \eta_{i}}(\xi) \,\la_{z\si}(d\xi)\,\la_{z\si}(d\ga) \\
&= z^{\sum\limits_{i \in I} \vert \eta_{i} \vert}\int_{\Ga_0}F( \sum_{i \in I} \eta_{i} +\ga)\,\la_{z\si}(d\ga).
\end{split}
\end{equation*}

\section{Correlation functions.}
\label{correl}
\setcounter{equation}{0} \renewcommand{\theequation}{\arabic{section}.%
\arabic{equation}} 

\subsection{Interaction potential.}

\begin{figure}[H]
\centering
\includegraphics[width=15cm,height=15cm,keepaspectratio]{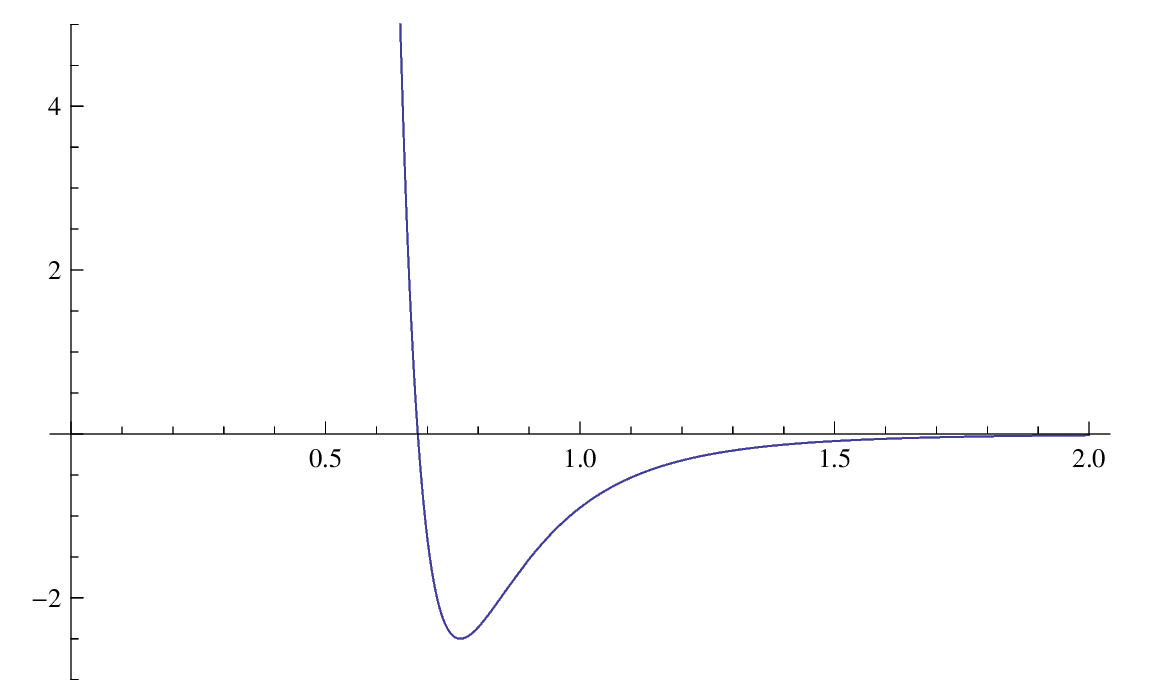}
\caption{The Lennard-Jones potential.}
\label{fig1}
\end{figure}

We consider a general type of two-body interaction potential
\begin{equation*}
V_2(x,y)=\phi(\vert x-y \vert),
\end{equation*}
where $\phi:[0,+\infty) \rightarrow \mathbb{R}\cup\{+\infty\}$ satisfies the
following conditions.\\
\indent \textbf{\small{(A): Assumptions about the interaction potential.}} \textit{The potential $\phi$\, is
continuous on $(0,+\infty)$, $\phi(0) = +\infty$,  and  there exist constants
$ 0 < r_1 < r_0 < r_2$, $\varphi_{1}>0$, $\varphi_{2}>0$, $s \geq d$ and $\varepsilon_0 > 0$ such that}
\begin{gather}
\phi(r) = \phi^+(r)\,\,\,\text{for}\,\,\,0 < r \leq r_0,\,\,\,\text{and}\,\,\,\phi^+(r)\geq \varphi_{1} r^{-s}\,\,\,
\text{for}\,\,\, r < r_1; \label{bihzero} \\  
\phi(r) = -\phi^-(r)\,\,\,\text{for}\,\,\,r > r_0,\,\,\,\text{and}\,\,\,\phi^-(r)\leq
\varphi_{2} r^{-d-\varepsilon_0}\,\,\,
  \text{for}\,\,\, r > r_2,\label{bihinfty}   
\end{gather}
where $\phi^{+}$ and $\phi^{-}$ denote the positive and negative parts of $\phi$ respectively defined as
\begin{equation*}
\phi^+(r):= \max \{0, \phi(r)\},\quad \phi^-(r):= -\min \{0,\phi(r)\}.
\end{equation*}
\noindent The shape of such potentials is illustrated in Figure \ref{fig1}.

A typical example is the Lennard-Jones potential, see, e.g., \cite{Ru69,Do99}, given by
\begin{equation*}
\phi_{\mathcal{LJ}}(\vert x \vert):= \frac{\varphi_0}{\vert x\vert^{6}} (\frac{r_0^6}{\vert x\vert^{6}}- 1),
\end{equation*}
where $\varphi_0>0$ is a given constant. It is clear that the potential $\phi_{\mathcal{LJ}}$ is strongly superstable, see, e.g., \cite{RT08}.\\
\indent Given $\eta, \ga \in\Ga_0$, we define the total particle interaction energy $U(\ga)$ in the configuration $\ga$ and the interaction energy $W(\eta; \gamma)$ between the particles in the configurations $\eta$ and $\ga$ respectively as
\begin{gather}
\label{U}
U(\gamma)=U_{\phi}(\gamma) := \sum_{\substack{\eta \leq \ga\\ \vert \eta\vert=2}} V_2(\eta),\\
\label{W}
W(\eta; \gamma):=\sum_{\substack{x\in\eta \\ y\in\gamma}} \eta(x) \ga(y) \phi(|x-y|).
\end{gather}
Note that, under our conditions, $U(\ga) = +\infty$ if $\ga(x) \geq 2$ for some $x$, and similarly, $W(\eta; \ga) = +\infty$ if $\eta$ and $\ga$ overlap, i.e. there exist some $x$ such that $\eta(x) \neq 0$ and $\ga(x) \neq 0$.

\begin{remark}
The conditions \eqref{bihzero} and \eqref{bihinfty}  are more restrictive than needed to obtain the basic expansions for the correlation functions. Sufficient assumptions to obtain analytic expansions are stability
\begin{equation}
\label{stU}
U(\gamma)\geq-B\vert \gamma\vert,\quad B \geq 0,\,\, \gamma\in\Gamma_0,
\end{equation}
and regularity, see, e.g., \cite[Sec. 4.1]{Ru69},
\begin{equation}
\label{reg}
\nu_{1}(\beta):= \int_{\R^d} \nu_{\beta}(x)\,dx < +\infty,\quad\textrm{with}\quad \nu_{\beta}(x) := \vert \mathrm{e}^{-\beta\phi(\vert x \vert)}-1\vert.
\end{equation}
We emphasize that \eqref{stU} and \eqref{reg} hold true under the conditions \eqref{bihzero} and \eqref{bihinfty}.
\end{remark}

\subsection{Gibbs measure.}

With the notation introduced above, given $\Lambda \in \mathcal{B}_{c}(\mathbb{R}^{d})$, the \textit{Gibbs measure} $\mu_{\La}$ on the configuration space $\Gamma_{\Lambda}$ is defined as
\begin{align}
\label{GMGCA}
\mu_{\La}(d\ga):={}&\frac{1}{Z_\La}\mathrm{e}^{-\be U(\ga)}\,\la_{z\sigma}(d\ga),\\
\label{GPF}
Z_\La:={}&\int_{\Ga_\La} \mathrm{e}^{-\be U(\ga)}\,\la_{z\sigma}(d\ga),
\end{align}
where $Z_{\Lambda}$ is the finite-volume partition function. For a survey and discussion of problems related to the construction of limit Gibbs measures for infinite systems in the space $\Ga$, we refer the readers to the review \cite{KPR12} and references therein.

\subsection{Correlation measure and correlation functions.}

\indent Correlation functions are the analogue of the moments of a measure. Let $\mathcal{M}^+(\R^d)$ denote the space of nonnegative Radon measures in $\mathcal{B}(\R^d)$. Consider the moments of a measure in the configuration space $\Ga$. With every configuration $\ga\in\Ga$ can be associated an occupation measure
according to, see, e.g., \cite{AKR98-1,Ito88},
\begin{equation*}
\Ga \ni \ga\mapsto\sum_{x\in\ga}\delta_x \in \mathcal{M}^+(\R^d),
\end{equation*}
where, as previously, $x \in \gamma$ means $x \in \mathbb{R}^{d}$ such that $\gamma(x) \geq 1$, and, $\delta_x$ is the Dirac measure, i.e.,
\begin{equation*}
\label{DF2}
\langle \delta_x, f \rangle =f(x),\quad f\in C_0(\R^d),
\end{equation*}
and $C_0(\R^d)$ denotes the space of continuous functions with compact support in $\mathbb{R}^{d}$.\\
\indent To generalize this to the case of several variables, note that the product of distributions is not defined. For example, in the case of Gaussian measures, one usually applies Wick regularization, see, e.g., \cite{Simon74, BK88}. An analogous procedure may be used for Poisson variables and is described below.
Note that due to our assumption $U(\gamma) = +\infty$ if $\ga(x)>1$ for some $x\in \mathbb{R}^{d}$, see \eqref{U}, we can restrict ourselves to configurations such that $\ga \leq 1$ from now on. The corresponding point process is then simple.\\
\indent Let $ F: \Ga_0 \rightarrow \R$ be a function on the configuration space $\Ga_0$  such that
\begin{equation*}
F\restriction\Ga^{(n)}:= F^{(n)}(\{x_1,\dots,x_n\})=F_n(x_1,\dots,x_n), \quad n \in \mathbb{N},
\end{equation*}
where, for every $n \in \mathbb{N}$, $F_n\in C_0(\R^{dn})$ is a symmetric function. Then,
\begin{equation*}
\langle F^{(1)},\ga \rangle := \sum_{x_1\in\ga} \langle F^{(1)},\delta_{x_1} \rangle  = \sum_{x_1\in\ga} F_1(x_1),
\end{equation*}
and we define the  $n$-th power by
\begin{equation}
\label{pwn}
\langle F^{(n)},:\ga^{\otimes n}: \rangle := \sum_{\substack{x_{1},\dots,x_{n} \in \mathbb{R}^{d}\\ \ga_{(x_1,\dots,x_n)} \leq \ga}} F_n(x_1,\dots,x_n).
\end{equation}
The \textit{correlation measures} $\rho^{(n)}$ are defined by
\begin{equation}
\label{mom}
\int_{\Ga} \langle F^{(n)},:\ga^{\otimes n}: \rangle \,\mu(d\ga) = \int_{\R^{dn}} F_n(x_1,\dots,x_n)\,\rho^{(n)}(dx_1,\dots,dx_n).
\end{equation}
In case that the correlation measures $\rho^{(n)}$ are absolutely continuous with respect to the Lebesgue measure in $\R^{dn}$, \textit{correlation functions} are defined as
\begin{equation*}
\rho^{(n)}(dx_1,\dots,dx_n) := \frac{1}{n!}\rho_n(x_1,\dots,x_n)\,dx_1 \cdots dx_n.
\end{equation*}
These functions are obviously symmetric, so that we can write
\begin{equation*}
\rho_n(x_1,\dots,x_n) = \rho(\eta)\restriction\Ga^{(n)},\quad \eta= \{x_1,\dots,x_n\}.
\end{equation*}
Using \eqref{pwn}, \eqref{mom} can then be rewritten in the form
\begin{equation*}
\int_{\Ga} \sum_{\substack{x_{1},\dots,x_{n} \in \mathbb{R}^{d}\\ \ga_{(x_1,\dots,x_n)} \leq \ga}} F_n(x_1,\dots,x_n)\,\mu(d\ga) = \int_{\R^{dn}} F_n(x_1,\dots,x_n)\,\rho^{(n)}(dx_1,\dots,dx_n).
\end{equation*}
Given $\La \in {\cal B}_c(\R^d)$, we can now define the \textit{correlation measure} $\rho$ on the configuration space $\Ga_\La$ by
\begin{equation*}
\int_{\Ga_\La} F(\eta)\,\rho(d\eta) = \sum_{n=0}^\infty \int_{\Ga_\La} \sum_{\substack{x_{1},\dots,x_{n} \in \mathbb{R}^{d}\\ \ga_{(x_1,\dots,x_n)} \leq \ga}}  F_n(x_1,\dots,x_n)\,\mu(d\ga),
\end{equation*}
where the term $n=0$ in the sum is set to $1$ by convention. In the case that the correlation measures are absolutely continuous, we have,
\begin{equation}
\label{opK-2}
\int_{\Ga_\La} F(\eta)\rho(\eta)\, \la_{\si}(d\eta) = \int_{\Ga_\La} \sum_{\eta \leq \ga} F(\eta)\, \mu(d\ga).
 \end{equation}
\indent From \eqref{opK-2} along with \eqref{GMGCA} and \eqref{equ}, the finite-volume correlation functions can be written as
\begin{equation}
\label{corf3}
\rho_\La(\eta)=\frac{z^{\vert \eta \vert}}{Z_\La} \int_{\Ga_\La} \mathrm{e}^{-\beta U(\eta + \ga)}\,\la_{z\si}(d\ga),\quad \eta \in \Gamma_{\Lambda}.
\end{equation}
Notice that $\rho_{\Lambda}(\eta) = 1$ whenever $\eta \in \Gamma_{\Lambda}^{(0)}$. Problems related to the construction of correlation functions in the infinite-volume limit are discussed in, e.g., \cite{Ru69, BPK69, Ru70, Re98, PR07, PR09}.

\subsection{Truncated (connected) correlation functions.}

Correlations between particles are better described by \textit{truncated (connected) correlation functions} (TCF).
Given $\eta \in \Gamma_{0}$ with $\vert \eta \vert=n\in \mathbb{N}$, these functions are defined recursively by
\begin{equation}
\begin{split}
\label{FunctT}
&\rho^T(x_1) := \rho(\mathbf{1}_{\{x_1\}}), \\
&\rho^T(x_1,\dots,x_n):=\rho(\eta)-\sum_{k=2}^{n} \sum_{I_1,\dots,I_k \subset \{1,\dots,n\}}^{*} \prod_{l=1}^k \rho^T(\eta_{I_l}),\quad n \geq 2,
\end{split}
\end{equation}
where $\rho(\eta)$ are the correlation functions, and the asterisk over the sum means that the sum is over all partitions of the set $\{1,\dots,n\}$ into $k$ non-empty disjoint subsets, and $\eta_{I} = \ga_{x_I}$. That is,
\begin{equation}\label{sum-cond}
\sum_{l=1}^k \eta_{I_l} = \ga_{(x_1,\dots,x_n)},\,\,\, \textrm{with $I_l \neq \emptyset$\, and \,$I_i \cap I_{i'} = \emptyset$\, if\, $i\neq i'$}.
\end{equation}
The TCF can also be written explicitly in terms of the correlation functions $\rho(\eta)$ as follows
\begin{equation}\label{FunctT1}
 \rho^T(x_1,\dots,x_n)=\sum\limits_{k=1}^{n}(-1)^{k-1}(k-1)! \sum_{
I_1,\dots,I_k \subset \{1,\dots,n\}}^{*} \prod_{l=1}^k \rho(\eta_{I_l}).
\end{equation}
Note that \eqref{FunctT1} is related to he M\"obius inversion formula, see, e.g., \cite[Chap. II]{MM85}.
Clearly, the TCF are permutation-invariant. They can then be written as $\rho^T(\ga_{(x_1,\dots,x_n)})$. In case that $\rho_{\Lambda}$ is given by \eqref{corf3}, the TCF have, in the thermodynamic limit, the following representation in terms of integrals with respect to the measure $\lambda_{z\sigma}$.

\begin{proposition}
\label{trunccor}
Assume that the interaction potential $\phi$ satisfies \eqref{stU} and \eqref{reg}.
Then, for all $\beta > 0$ and for all $0<z< r(\beta)$ with,
\begin{equation}
\label{rbeta}
r(\beta) := \mathrm{e}^{-2\beta B-1}\nu_{1}(\beta)^{-1},
\end{equation}
the following representation for the TCF holds true
\begin{equation}
\label{trunccoreq}
\rho^T(\eta)=z^{|\eta|}\int_{\Gamma_{0}} \Phi^T(\eta + \gamma)\, \lambda_{z\sigma}(d\gamma),\quad \eta \in \Gamma_{0}.
\end{equation}
Here, the function $\Phi^T(\gamma)$ is the so-called Ursell function, see, e.g., \cite{Ru69}, given by
\begin{equation*}
  \Phi^T(\ga)
 = \left\{\begin{array}{ll}
    0,   &\mbox{if $\ga=0,$} \\
    1,   &\mbox{if $|\ga|= 1,$} \\
    \sum\limits_{G\in\mathcal{G}^T(\ga)} \prod\limits_{\{x,y\}\in\mathcal{L}(G)} C_{xy}, &\mbox{if $|\ga|\geq 2$,}
 \end{array}
 \right.
\end{equation*}
in which $\mathcal{G}^T(\ga)$ stands for the set of all connected graphs $G$ (Mayer graphs) with vertices in the points $x$ of the configuration $\ga$, and $\mathcal{L}(G)$ is the set of all edges of the graph $G$, and
\begin{equation}
\label{Cxy}
C_{xy}:=\mathrm{e}^{-\beta\phi(|x-y|)}-1.
\end{equation}
Moreover, the TCF in \eqref{trunccoreq} can be analytically extended to the open disk of radius $r(\beta)$ given in \eqref{rbeta}.
\end{proposition}

For a proof, we refer the readers to \cite{Pe63,Ru64}. See also \cite{Pe67} and \cite[Sec. 4]{Ru69}.\\
\noindent In his proof \cite{Pe63}, O. Penrose noted that one could associate with each connected graph $G$ on $\gamma$ a unique Cayley tree obtained by deleting bonds from $G$ in a particular way (tree graph identity). The sum over connected graphs may be rearranged by grouping together all terms (graph contributions) corresponding to a given Cayley tree, which are obtained by the procedure of "deleting". Later, D. Brydges and P. Federbush proposed in \cite{BF78} a new method to derive the Mayer series for the pressure via a new type of tree graph identity. A more detailed history of the subject and some new results can be found in \cite{MPS14}.\\
\indent In this article, we derive an expansion for more general PTCF using the technique of R.A. Minlos and S.K. Poghosyan in \cite{MP77} (see also \cite{PU09}) which is related to Penrose's original proof.
A representation for the functions $\rho^T$ in the form of expansions in terms of contributions
from tree graphs follows as a special case.

\subsection{Partially truncated (connected) correlation functions}

\textit{Partially truncated (connected) correlation functions} (PTCF)  describe correlations between clusters of particles. Decay estimates for these correlations are an important technical tool in the proof of many physical hypotheses. For instance, see \cite[Eq. (4.2)]{BKL85}.\\
\indent Given $m \in \mathbb{N}$, consider a collection $(\eta_{i})_{i=1}^{m}$ of configurations $\eta_{i} \in \Gamma_{0}$ (for instance, resulting from the decomposition of a given $\overline{\eta} \in \Gamma_{0}$ into $m$ clusters). The corresponding PTCF are defined recursively by
\begin{equation}
\label{FunctPT}
\begin{split}
&\w\rho^T(\eta_1):= \rho(\eta_1), \\
&\w\rho^T(\eta_1;\dots;\eta_m):=\rho(\sum_{i=1}^m \eta_i)-\sum_{k=2}^{m} \sum_{I_1,\dots,I_k \subset \{1,\dots,m\}}^{*} \prod_{l=1}^{k} \w\rho^T(\w\eta_l),\quad m \geq 2,
\end{split}
\end{equation}
where, as previously, the asterisk over the sum means that the sum is over all partitions of $\{1,\dots,m\}$ into $k$ non-empty disjoint subsets, and where
\begin{equation*}
\w\eta_l := \sum_{i \in I_l} \eta_i \quad \textrm{and} \quad \sum_{l=1}^k \w\eta_l = \sum_{i=1}^m \eta_i.
\end{equation*}
We will sometimes use the notation $\w\rho_m^T(\eta_1;\ldots;\eta_m)=\w\rho^T(\eta_1;\dots;\eta_m)$ to emphasize the number of clusters. Obviously, definition \eqref{FunctPT} coincides with the TCF in \eqref{FunctT} when all configurations $\eta_i$ consist of exactly one point. Analogous to \eqref{FunctT1}, the PTCF can be expressed directly in terms of the $\rho(\tilde{\eta}_i)$ as
\begin{equation}
\label{FunctPT2}
\tilde{\rho}^T(\eta_1;\dots;\eta_m) = \sum_{k=1}^m (-1)^{k-1} (k-1)! \sum_{I_1, \dots,I_k \subset \{1,\dots,m\}}^{*}\prod_{l=1}^{k} \rho(\tilde{\eta}_l).
\end{equation}
\indent To derive such an expression for the PTCF, we introduce a generating functional. It is a generalization of the generating functional introduced in \cite{IS78} for spin systems.\\
\indent For a given nonnegative $j\in C^\infty_0(\R^d)$, define the smoothed correlation function $\rho_{j}$ by
\begin{equation}
\label{FunctSF}
\rho_{j}(\eta)=\rho_{j;1}(\eta) := \frac{z^{\vert \eta \vert}}{Z_j} \int_{\Ga_0} \chi_j(\eta + \ga) \mathrm{e}^{-\beta U(\eta + \ga)} \,\la_{z\si}(d\ga),\quad \eta \in \Gamma_{0},
\end{equation}
where the function $\chi_{j}: \Gamma_{0} \rightarrow \mathbb{R}_{+}$ is defined as in \eqref{LPE}, and
\begin{equation*}
Z_j := \int_{\Ga_0} \chi_j(\ga) \mathrm{e}^{-\beta U(\ga)} \,\la_{z\si}(d\ga).
\end{equation*}
Using the definitions \eqref{DF} and \eqref{prd1} with nonnegative reals $(\alpha_{i})_{i=1}^{m}$, we now put
\begin{equation}
\label{FunctG}
\w{F}^T_{\rho_j}(\alpha,\eta)_1^m := \log (Z_j((\alpha_{i},\eta_{i})_{i=1}^{m})),
\end{equation}
where
\begin{equation}
\label{FunctPF}
Z_j((\alpha_{i},\eta_{i})_{i=1}^{m}) := \langle \prod_{i=1}^m \Delta_{(\alpha_i,\eta_i)}, \chi_j  \mathrm{e}^{-\beta U} \rangle = \int_{\Gamma_0}\prod_{i=1}^m\Delta_{(\alpha_i\eta_i)}(\gamma) \chi_j(\gamma)  \mathrm{e}^{-\be U(\ga)}\,\la_{z\sigma}(d\ga).
\end{equation}
Here \eqref{FunctPF} stands for a mixed partition function for a gas together with $m$ clusters $\eta_1,\dots,\eta_m$ having additional activity parameters $\alpha_i$. We note that the use of smooth cutoff functions $j$ (instead of indicator functions) ensures that the quantity in \eqref{FunctPF} is well-defined, see Sec. \ref{rsecto}. In case $\alpha_i=0$, $i=1,\dots,m$ we can put $j(x)=\mathbf{1}_\Lambda(x)$, where $\mathbf{1}_\Lambda$ denotes the indicator function of a set $\Lambda \in \mathcal{B}_c(\mathbb{R}^{d})$ and \eqref{FunctPF} reduces to the partition function in \eqref{GPF}. Further, define
\begin{equation}
\label{FunctHG}
\w\rho^T_{j;r}(\eta_1;\dots;\eta_r\vert (\alpha_{i},\eta_{i})_{i=r+1}^{m}):= (\prod_{i=1}^{r}\frac{\partial}{\partial \alpha_i})
\w{F}^T_{\rho_j}((\alpha_{i},\eta_{i})_{i=1}^{m})\big|_{\alpha_1=\cdots=\alpha_r=0},\quad 1\leq r\leq m.
\end{equation}
We call \textit{$r$-point $j$-PTCF}, or simply \textit{$j$-PTCF} when $r=m$, the following functions
\begin{equation}
\label{FunctFG}
\w\rho^T_{j;r}(\eta_1;\dots;\eta_r):=\w\rho^T_{j;r}(\eta_1;\dots;\eta_r\vert (\alpha_{i},\eta_{i})_{i=r+1}^{m})\big|_{\alpha_{r+1}=\cdots=\alpha_m=0}.
\end{equation}

We conclude this section with the following lemma

\begin{lemma}
Given $r,m \in \mathbb{N}$ such that $1 \leq r \leq m$, the $r$-point $j$-PTCF associated to the collection $(\eta_{i})_{i=1}^{m}$ of configurations $\eta_{i}\in \Gamma_{0}$ are given by
\begin{equation}
\label{PTCF}
\w\rho^T_{j;r}(\eta_1; \dots; \eta_r) = \sum_{k=1}^r (-1)^{k-1} (k-1)! \sum_{\{J_1,\dots,J_k\} \subset \{1,\dots,r\}}^{*} \prod_{l=1}^k \rho_j (\sum_{i\in J_l} \eta_i),
\end{equation}
where the second sum in \eqref{PTCF} runs over all partitions of $\{1,\dots,r\}$ into $k$ non-empty subsets $J_1,\dots,J_k$ with the restrictions \eqref{sum-cond}. In particular, when $j(x)=\mathbf{1}_\Lambda(x)$ the functions \eqref{FunctFG} correspond to the finite-volume PTCF in $\Lambda$ and when $j(x)=1$ they correspond to the PTCF in $\R^d$ given in \eqref{FunctPT2}.
\end{lemma}

\begin{remark}
\label{rem}
One can show by induction that, for $r\geq 2$, $\w\rho^T_{j;r}(\eta_1; \dots; \eta_r)=0$ if there exists $i_{0} \in \{1,\dots,r\}$ such that $\vert \eta_{i_{0}}\vert=0$, see \eqref{PTCF} along with \eqref{FunctSF}.
\end{remark}

\noindent \textbf{Proof.} The key ingredient is the following formula. Given a smooth function $Z: \mathbb{R}^{m} \rightarrow (0,+\infty)$,
\begin{multline}
\label{LogZder}
(\prod_{i=1}^{k}\frac{\partial}{\partial \alpha_i}) \log (Z((\alpha_{i})_{i=1}^{m})) = \\ \sum_{k=1}^r (-1)^{k-1}(k-1)! \sum_{\{J_1,\dots,J_k\} \subset \{1,\dots,r\}}^{*} \prod_{l=1}^k \frac{1}{Z((\alpha_{i})_{i=1}^{m})} (\prod_{i\in J_l} \frac{\partial}{\partial \alpha_i})Z((\alpha_{i})_{i=1}^{m}),\quad 1\leq r \leq m,
\end{multline}
which easily follows by induction. On the other hand, from \eqref{FunctPF} along with \eqref{Deltadistr}, we have,
\begin{multline*}
(\prod_{i\in J_l} \frac{\partial}{\partial \alpha_i})Z_j((\alpha_{i},\eta_{i})_{i=1}^{m}) = \\ z^{\sum\limits_{i\in J_{l}} |\eta_i|} \sum_{I \subset \{1,\dots,m\}\setminus J_{l}}
\prod_{i\in I} (\alpha_i z^{|\eta_i|}) \int_{\Gamma_0} \chi_j(\sum_{i \in J_{l}} \eta_i + \sum_{i \in I} \eta_i + \ga) \mathrm{e}^{-\beta U(\sum\limits_{i \in J_{l}} \eta_i + \sum\limits_{i \in I} \eta_i + \ga)}\,\lambda_{z\si}(d\ga).
\end{multline*}
Setting the remaining $\alpha_i = 0$, only the empty set $I=\emptyset$ survives. In view of \eqref{FunctSF}, we then obtain
\begin{equation}
\label{id11} 
\frac{1}{Z_j((\alpha_{i},\eta_{i})_{i=1}^{m})} (\prod_{i\in J_l} \frac{\partial}{\partial \alpha_i}) Z_j((\alpha_{i},\eta_{i})_{i=1}^{m}) \big\vert_{\alpha_{1} = \cdots \alpha_{m} = 0} =
\rho_j (\sum_{i \in J_{l}} \eta_i).
\end{equation}
Replacing $Z((\alpha_{i})_{i=1}^{m})$ by $Z_j((\alpha_{i},\eta_{i})_{i=1}^{m})$ in \eqref{LogZder}, \eqref{PTCF} follows from \eqref{id11}. \hfill $\blacksquare$ \\
\indent In particular, taking the limit $j \to 1$ in \eqref{PTCF} with $k=m$, we obtain \eqref{FunctPT2}.

\section{Equations for PTCF and their solutions.}
\setcounter{equation}{0} \renewcommand{\theequation}{\arabic{section}.%
\arabic{equation}} 

\subsection{Kirkwood-Salsburg-type equations.}

Let $(\eta_{i})_{i=1}^{m}$, $m\geq 2$ be a collection of configurations in $ \Gamma_{0}$ such that $\sum_{i=1}^{m} \vert \eta_{i} \vert>0$. We start by deriving Kirkwood-Salsburg-type equations for the 1-point $j$-PTCF. Assume that $\vert \eta_{1}\vert>0$ (we may change the cluster labelling if needed). From \eqref{FunctFG}--\eqref{FunctHG} (with $k=1$) and \eqref{FunctG}--\eqref{FunctPF}, the 1-point $j$-PTCF reads
\begin{equation}
\label{1ptPTCF}
\tilde{\rho}^{T}_{j;1}(\eta_{1}\vert(\alpha_{i},\eta_{i})_{i=2}^{m}) = \frac{1}{Z_{j}((\alpha_{i},\eta_{i})_{i=1}^{m})} \frac{\partial}{\partial \alpha_{1}} \int_{\Gamma_0}\prod_{i=1}^m\Delta_{(\alpha_i\eta_i)}(\gamma) \chi_j(\gamma)  \mathrm{e}^{-\be U(\ga)}\,\la_{z\sigma}(d\ga)\big\vert_{\alpha_{1}=0}.
\end{equation}
In view of \eqref{W}, let $x_1\in\eta_1$ such that writing
\begin{equation*}
\eta_{1}^{'} := \eta_1- \mathbf{1}_{\{x_1\}},
\end{equation*}
we have,
\begin{equation*}
W(x_1;\eta^{'}_1) := W(\mathbf{1}_{\{x_1\}};\eta_{1}^{'}) \geq -2B,
\end{equation*}
where $B \geq 0$ is defined by \eqref{stU}. Note that the existence of such a point in any configuration follows from \eqref{stU}, see \cite[Chap. 4]{Ru69}. Note also that $W(x_1;\eta^{'}_1) =+\infty$ if $\eta_{1}(x_1) >1$. Consider the decomposition,
\begin{equation}
\label{FunctHG21}
\mathrm{e}^{-\be U(\eta_1+\ga)} = \mathrm{e}^{-\beta W(x_1;\eta^{'}_1)} \mathrm{e}^{-\beta W(x_1;\gamma)} \mathrm{e}^{-\beta U(\eta^{'}_1 + \ga)} = \mathrm{e}^{-\beta W(x_1;\eta^{'}_1)} \sum_{\xi \leq \gamma} K(x_1;\xi) \mathrm{e}^{-\beta U(\eta^{'}_1  + \ga)},\quad \gamma \in \Gamma_{0},
\end{equation}
where
\begin{equation*}
K(x_1;\xi):= \prod_{y\in\xi} C_{x_1 y} = \prod_{y\in\xi} (\mathrm{e}^{-\beta\phi(\vert x_1-y \vert)}-1).
\end{equation*}
Inserting the right-hand side of the second equality of \eqref{FunctHG21} into \eqref{1ptPTCF}, we have,
\begin{equation}
\label{FunctHG22}
\w\rho^T_{j;1}(\eta_1\vert(\alpha_{i},\eta_{i})_{i=2}^m) = \frac{z^{\vert \eta_1 \vert} \mathrm{e}^{-\beta W(x_1;\eta^{'}_1)}}{Z_j((\alpha_{i},\eta_{i})_{i=2}^m)} \int_{\Gamma_0} \sum_{\xi\leq \gamma} K(x_1;\xi) \prod_{i=2}^m \Delta_{(\alpha_i\eta_i)}(\gamma) \chi_j(\eta_1 + \gamma) \mathrm{e}^{-\beta U(\eta^{'}_1 + \ga)}\, \la_{z\sigma}(d\ga).
\end{equation}
Putting $\alpha_2=\dots=\alpha_m=0$ in \eqref{FunctHG22} and then using \eqref{equ} (extended to the configuration space $\Gamma_{0}$) along with the identity $\w\rho^T_{j;1}(\eta^{'}_1 + \ga) = \rho_j(\eta^{'}_1 + \ga)$, we obtain the Kirkwood-Salsburg equation
\begin{equation}
\label{KS1}
\w\rho^T_{j;1}(\eta_1) = \rho_j(\eta_1) = z \mathrm{e}^{-\beta W(x_1;\eta^{'}_1)} j(x_1)\int_{\Gamma_{0}} K(x_1;\xi) \rho_j(\eta^{'}_1 + \xi)\,\la_{\sigma}(d\xi).
\end{equation}
We now generalize those equations for the $m$-point $j$-PTCF as follows. \eqref{KS1} can be generalized as
\begin{equation}
\label{KS11} 
\rho_j(\eta_1 + \eta) = z \mathrm{e}^{-\beta W(x_1;\eta^{'}_1)} j(x_1)
\sum_{\xi \leq \eta} \int_{\Gamma_0} K(x_1;\xi + \ga) \rho_j(\eta^{'}_1 + \eta + \ga)\,\la_{\sigma}(d\ga),
\end{equation}
where we used in the expansion \eqref{FunctHG21} the identity
\begin{equation}
\label{Ident} 
\sum_{\xi \leq \gamma+\eta}K(x_1;\xi) = \sum_{\xi \leq \eta}\sum_{\varsigma \leq \gamma}K(x_1;\xi+\varsigma).
\end{equation}
Inserting \eqref{KS11} into \eqref{PTCF} (instead of $\rho_j$ which contains $\eta_1$) and denoting $I_1 := J_1\setminus\{1\}$, we have
\begin{multline*}
\w\rho^T_{j;m}(\eta_1;\dots;\eta_m) = z \mathrm{e}^{-\beta W(x_1;\eta^{'}_1)} j(x_1) \sum_{k=1}^m (-1)^{k-1} (k-1)! \\
\times \sum_{I_1 \subset \{2,\dots,m\}} \sum_{\{I_2,\dots,I_k\} \subset \{2,\dots,m\}\setminus I_1}^{*}
\sum_{\xi \leq \sum\limits_{i \in I_1} \eta_i} \int_{\Ga_0} K(x_1;\xi + \ga)
\rho (\eta^{'}_1 + \sum_{i \in I_1} \eta_i + \ga) \prod_{l=2}^k \rho (\sum\limits_{i \in I_l} \eta_i)\, \la_\si(d\ga).
\end{multline*}
Note that, in the second sum, the set $I_1$ can take on the value $I_1=\emptyset$ in contrast to
$I_2,\dots, I_k$ in the 3-d sum. Changing the order of summations over indices $I$ and over sets $\xi$, we may write
\begin{multline*}
\w\rho^T_{j;m}(\eta_1; \dots; \eta_m) = z \mathrm{e}^{-\beta W(x_1;\eta^{'}_1)} j(x_1)
\sum_{\xi \leq \sum\limits_{i \in I_1} \eta_i} \int_{\Ga_0}\la_\si(d\ga)\, K(x_1;\xi + \ga)
\sum_{k=1}^{m-\vert I_0(\xi)\vert} (-1)^{k-1} (k-1)! \\
\times \sum_{I_1 \subset \{2,\dots,m\}} \sum_{\{I_2,\dots,I_k\} \subset \{2,\dots,m\}\setminus (I_0(\xi)\cup I_1 )}^{*} \rho (\eta^{'}_1 + \sum_{i \in (I_{0}(\xi) \cup I_1)} \eta_i + \ga)
\prod_{l=2}^k \rho (\sum\limits_{i \in I_l} \eta_i),
\end{multline*}
where we set $I_0(\xi):= \{i \geq 2\,:\, \xi \cdot \eta_i \neq 0\}$. Setting now $\eta_{\{2,\dots,m\}\setminus I} := (\eta_{i_2}; \dots;\eta_{i_{m-|I|}})$ if $\{2,\dots,m\} \setminus I = \{i_2,\dots,i_{m-|I|}\}$, we arrive at
\begin{multline*}
\w\rho^T_{j;m}(\eta_1; \dots; \eta_m) = \\ z \mathrm{e}^{-\beta W(x_1;\eta^{'}_1)} j(x_1) \sum_{\xi \leq \sum\limits_{i=2}^m \eta_i} \int_\Ga  K(x_1;\xi + \ga)
\w\rho^T_{m-\vert I_0(\xi)\vert}(\eta^{'}_1 + \sum_{i \in I_0(\xi)}\eta_i + \ga; \eta_{\{2,\dots,m\} \setminus I_0(\xi)})\,\la_\si(d\ga).
\end{multline*}
Here is the final rewriting for the key equation (recursion relation)
\begin{multline}
\label{FunctPTCmnew}
\w\rho^T_{j;m}(\eta_1;\dots;\eta_m) = z \mathrm{e}^{-\beta W(x_1;\eta^{'}_1)} j(x_1) \\
\times \sum_{I \subset\{2,\dots, m\}}\sum^{*}_{\xi \leq \sum\limits_{i \in I} \eta_{i}} \int_{\Gamma_0} K(x_1;\xi + \ga) \w\rho^T_{j;m-|I|} (\eta_1 - \mathbf{1}_{\{x_1\}} + \sum_{i \in I} \eta_i + \gamma;\eta_{\{2,\dots,m\} \setminus I})\,\la_\si(d\ga),
\end{multline}
where the asterisk over the second sum means that for all $i \in I$, $\xi \cdot \eta_i \neq 0$. We emphasize that these equations hold provided that $\vert \eta_1 \vert > 0$. They express the $m$-point $j$-PTCF $\w\rho^T_{j;m}$ in terms of the $r$-point $j$-PTCF $\w\rho^T_{j;r}$, $r \leq m$. They therefore determine $\w\rho^T_{j;m}$ uniquely if the operator
\begin{equation*}
f \mapsto z\,\mathrm{e}^{-\beta W(x_1;\eta^{'})} \chi_j(\eta)(1-\delta_0(\eta)) \int_{\Ga_0} K(x_1;\ga) f(\ga + \eta - \mathbf{1}_{\{x_1\}})\, \la_\si(d\ga),
\end{equation*}
has $L^1(\Ga_0,\la_\sigma)$-norm less than 1. Here, $\delta_0(\eta) := 1$ if $\vert \eta \vert =0$, $\delta_0(\eta) := 0$ otherwise.

\begin{remark}
We point out that \eqref{FunctPTCmnew} can be alternatively obtained by taking the derivatives of \eqref{FunctHG22} with respect to $\alpha_2,\dots,\alpha_m$, see \eqref{FunctHG}--\eqref{FunctFG}, and by using \eqref{PTCF} along with \eqref{Ident}.
\end{remark}

\subsection{Solution in the thermodynamic limit.}

We adopt set notation from now on and write $\ga$ for the set of points $x$ with $\ga(x) =1$.\\
Following the strategy used in \cite{MP77}, we seek a solution of the equation \eqref{FunctPTCmnew} in the form
\begin{equation}
\label{rhom}
\w\rho^T_{j;m}(\eta_1;\dots;\eta_m)=\int_{\Gamma_0} \chi_j(\bigcup_{i=1}^m \eta_i \cup \gamma) T_m(\eta_1;\dots;\eta_m\mid\gamma)\,\la_{\sigma}(d\gamma),
\end{equation}
where $T_m(\eta_1;\dots;\eta_m\mid\gamma)$, $m\geq 2$ and $\gamma \in \Gamma_{0}$ is a family of kernels such that
\begin{equation*}
T_m(\eta_1; \dots; \eta_m \mid \ga) =0\,\,\, \textrm{if $\ga \cap \overline{\eta}\neq \emptyset$},\quad \overline{\eta}:= \bigcup_{i=1}^m \eta_i.
\end{equation*}
\indent Inserting the expression \eqref{rhom} for $\w\rho^T_{j;m}$ and $\w\rho^T_{j;m-|I|}$ in both sides of \eqref{FunctPTCmnew} and then applying Lemma \ref{combresum} (extended to the configuration space $\Gamma_{0}$), we arrive at the following recursion relations for the kernels $T_m(\eta_1;\dots;\eta_m\mid\gamma)$, owing to the arbitrariness of the function $j$,
\begin{multline}
\label{Tkernel}
T_m(\eta_1;\dots;\eta_m\mid\gamma) = \\
z \mathrm{e}^{-\beta W(x_1;\eta^{'}_1)} \sum_{\xi\subset\gamma} \sum_{ I\subset\{2,\dots, m\}} \sum^{*}_{\eta \subseteq \ov\eta_I} K(x_1;\eta \cup \xi) T_{m-\vert I\vert} (\eta_1' \cup \ov\eta_I \cup \xi; \eta_{\{2,\dots,m\}\setminus I}\mid\gamma \setminus \xi),
\end{multline}
where $\eta_1' = \eta_1 \setminus\{x_1\}$ (set notation) and where we set $\ov{\eta}_{I} := \bigcup_{i \in I} \eta_i$. Subject to the initial conditions
\begin{equation}
\label{cond1}
T_1(\emptyset\mid\emptyset)=1,\quad T_1(\emptyset \mid \ga) = 0\,\,\textrm{if  $\ga \neq \emptyset$},
\end{equation}
and also, for all $m>1$,
\begin{equation}
\label{cond2}
T_m(\eta_1;\dots;\eta_m\mid\gamma)=0\,\,\textrm{if $\gamma \neq \emptyset$ and $\eta_i = \emptyset$ for some $i=1,\dots,m$};
\end{equation}
the equation \eqref{Tkernel} has a unique solution due to its recursive structure. Indeed, these conditions follow from the fact that $\rho_j(\emptyset) = 1$ and $\w\rho^T_{j;m}(\eta_1;\dots;\eta_m) = 0$ if $\eta_i = \emptyset$ for some $i=1,\dots,m$, see Remark \ref{rem}.\\
\indent The main result of this section is a uniqueness result in the infinite-volume limit (i.e., $j \to 1$)

\begin{theorem}
\label{exist}
Assume that the interaction potential $\phi$ satisfies \eqref{bihzero} and \eqref{bihinfty}. Given $m \in \mathbb{N}$, $m\geq 2$, there exists, for all $\beta>0$, a unique solution of the equation \eqref{FunctPTCmnew} in the thermodynamic limit $j\to 1$, which can be written in the form
\begin{equation}
\label{rhomj=1}
\w\rho^T_{m}(\eta_1;\dots;\eta_m) = \int_{\Gamma_0} T_m(\eta_1;\dots;\eta_m\mid\gamma)\, \la_{\sigma}(d\gamma),
\end{equation}
where the family of kernels $\vert T_m(\eta_1;\dots;\eta_m\mid\ga)\vert$, $\gamma \in \Gamma_{0}$ is bounded above by a power series in the activity $z$ (with integrable coefficients) which converges in the region
\begin{equation}
\label{rad}
z \mathrm{e}^{2\beta B+2}\nu_{1}(\beta) < 1.
\end{equation}
Here, $B\geq 0$ and $\nu_{1}(\beta)>0$ are respectively defined in \eqref{stU} and \eqref{reg}.
\end{theorem}

\begin{remark}
We point out that condition \eqref{rad} given in Theorem \ref{exist} is not optimal and can be replaced by the condition
\begin{equation*}
z \mathrm{e}^{2\beta B+1}\nu_{1}(\beta) < 1
\end{equation*}
as in the standard Kirkwood-Salsburg equation. However, choosing a smaller domain of convergence allows us to derive a simple expression for an estimate of \eqref{rhomj=1}, see \eqref{upbdKS} and Remark \ref{tocom}.
\end{remark}

The remaining of this paragraph is devoted to the proof of Theorem \ref{exist}. For reader's convenience, the proofs of the intermediate results are placed in Sec. \ref{prroof}. Note that, in order to prove that \eqref{rhom}--\eqref{Tkernel} with the conditions \eqref{cond1}--\eqref{cond2} is a solution of the equation \eqref{FunctPTCmnew} as $j\to 1$, it is necessary to show that the kernels $T_m(\eta_1;\dots;\eta_m\mid\gamma)$ are integrable functionals of the variable $\gamma$ with respect to the measure $\la_{\sigma}$.\\
\indent Following \cite{MP77}, given $h>0$ and any bounded nonnegative even function $\nu:\R^d \to [0,+\infty)$, introduce a new family of kernels $Q_m(\eta_1;\dots;\eta_m\mid\gamma)$, $m \geq 2$ and $\gamma \in \Gamma_{0}$ which are uniquely determined by the following system of recursion relations
\begin{equation}
\label{Qkernel}
Q_m(\eta_1;\dots;\eta_m\mid\gamma) =
h\sum_{\xi\subset\gamma} \sum_{ I\subset\{2,\dots, m\}} \sum^{*}_{\eta \subseteq \ov\eta_I} K_\nu(x_1;\eta\cup\xi) Q_{m-|I|}(\eta_1' \cup \ov\eta_I \cup \xi; \eta_{\{2,\dots,m\}\setminus I} \mid \gamma \setminus \xi),
\end{equation}
with initial conditions
\begin{equation}
\label{Qcond1}
Q_1(\emptyset\mid\emptyset) = 1,\quad Q_1(\emptyset \mid \ga) = 0\,\,\textrm{if  $\ga \neq \emptyset$},
\end{equation}
and also, for all $m>1$,
\begin{equation}
\label{Qcond2}
Q_m(\eta_1;\dots;\eta_m\mid\gamma)=0\,\,\textrm{if $\gamma \neq \emptyset$ and $\eta_i = \emptyset$ for some $i=1,\dots,m$},
\end{equation}
where
\begin{equation}
\label{nukern}
K_\nu(x_1;\xi) :=
\left\{ \begin{array}{ll} 1,  &\textrm{if $\xi =\emptyset$}, \\
\prod\limits_{x\in\xi}\nu(x_1-x),  &\textrm{if $\vert \xi \vert\geq 1$.}
\end{array}\right.
\end{equation}
\indent Since, by using assumption \eqref{stU}, we have from the expression \eqref{Tkernel}
\begin{equation*}
\vert T_m(\eta_1;\dots;\eta_m\mid\gamma)\vert
\leq z \mathrm{e}^{2\beta B} \sum_{\xi\subset\gamma} \sum_{ I\subset\{2,\dots, m\}}
\sum^{*}_{\eta  \subseteq \ov\eta_I} \vert K(x_1;\eta \cup \xi)\vert \vert T_{m-|I|}(\eta_1' \cup \ov{\eta}_{I} \cup \xi; \eta_{\{2,\dots,m\}\setminus I} \mid \gamma \setminus \xi)\vert,
\end{equation*}
the following result, which can be proven by induction, is straightforward

\begin{lemma}
\label{riglem}
Assume that the interaction potential $\phi$ satisfies \eqref{bihzero}--\eqref{bihinfty}. Given $\beta>0$ and $z>0$, set
\begin{equation}
\label{cond3}
z\mathrm{e}^{2\beta B}=h\quad \textrm{and} \quad \vert\mathrm{e}^{-\beta\phi(|x-y|)}-1\vert =\nu(x-y),
\end{equation}
where $B \geq 0$ is defined in \eqref{stU}. Then, given $m \in \mathbb{N}$,  $m \geq 2$ and $\gamma \in \Gamma_{0}$, the following holds,
\begin{equation}
\label{TQ}
\vert T_m(\eta_1;\dots;\eta_m\mid\gamma)\vert \leq Q_m(\eta_1;\dots;\eta_m\mid\gamma).
\end{equation}
\end{lemma}

The solution $Q_m(\eta_1;\dots;\eta_m\mid\gamma)$ of the equation \eqref{Qkernel} with conditions \eqref{Qcond1}--\eqref{nukern} can be written with the help of \textit{forest graphs}. For each set of clusters $\{\eta_1;\dots;\eta_m\}$ with $\eta_j\in (\Gamma_0\setminus\emptyset)$ and each configuration $\gamma\in\Gamma_0$, we define the set of forest graphs $\mathfrak{S}(\eta_1;\dots;\eta_m\mid\gamma)$ in the following way. The connected components of the graphs $\w{f}\in \mathfrak{S}(\eta_1;\dots;\eta_m\mid\gamma)$ are tree graphs with vertices given by points of $\bigcup_{i=1}^{m}\eta_i\cup\gamma$, and such that there are no edges (or lines) connecting vertices of the same cluster $\eta_i$ (for $i = 1,\dots,m$). Each tree contains a point of  $\bigcup_{i=1}^{m} \eta_{i}$ and, if $i_0$ is the lowest index such that $\eta_{i_0}$ contains a point of the tree, then this point is unique (the \textit{root} of the tree). Moreover, for every other vertex $z$ of the tree there is a path $z_1, \dots, z_k$ such that $z_k=z$, and there is an edge between the root $x_0$ and $z_1$ and between each pair $z_p$ and $z_{p+1}$, and such that if $z_p \in \eta_{i_p}$ then, if $z_{p+1} \in \eta_{i_{p+1}}$ then $z_{p}$ is the only point in $\eta_{i_p}$ connected to a point in $\eta_{i_{p+1}}$ by a edge \textit{in the forest}, whereas if $z_{p+1} \in \ga$ then $z_p$ is the only point in $\eta_{i_p}$ to which it is connected by a edge in the forest.
Note that a single point $x \in \bigcup_{i=1}^{m}\eta_i$ is also a tree with analytic contribution $h$.
Finally, if all points of the configurations $\eta_i$ (for every $i=1,\dots,m$) are combined into one single vertex (and corresponding edges into a single edge), then the forest graph $\w{f}\in \mathfrak{S}(\eta_1;\dots;\eta_m\mid\gamma)$ reduces to a  connected tree graph with $m+n$ vertices, where $n=\vert \gamma \vert$.\\
The figure \ref{fig2} below illustrates typical examples of forest graphs. The forest in the first graph consists of two trees; the first has a root in $\eta_1$, the second in $\eta_2$. Note that a single point in $\eta_i$ can be connected to several points in $\eta_j$, hence the factor $L_j$. Note also that for example only the first point of $\eta_2$ is connected to $\eta_3$. The forest in the second graph illustrates the fact that in the path $z_1,\dots,z_p$, if $z_p \in \eta_{i_p}$ and $z_{p+1} \in \eta_{i_{p+1}}$, then it is possible that $i_p > i_{p+1}$ provided the root is in $\eta_{i}$ with $i < i_{p+1}$. For example the second vertex of $\eta_2$ and the first vertex of $\eta_3$ with root in $\eta_1$.

\begin{figure}[H]
\centering
\includegraphics[width=15cm,height=15cm,keepaspectratio]{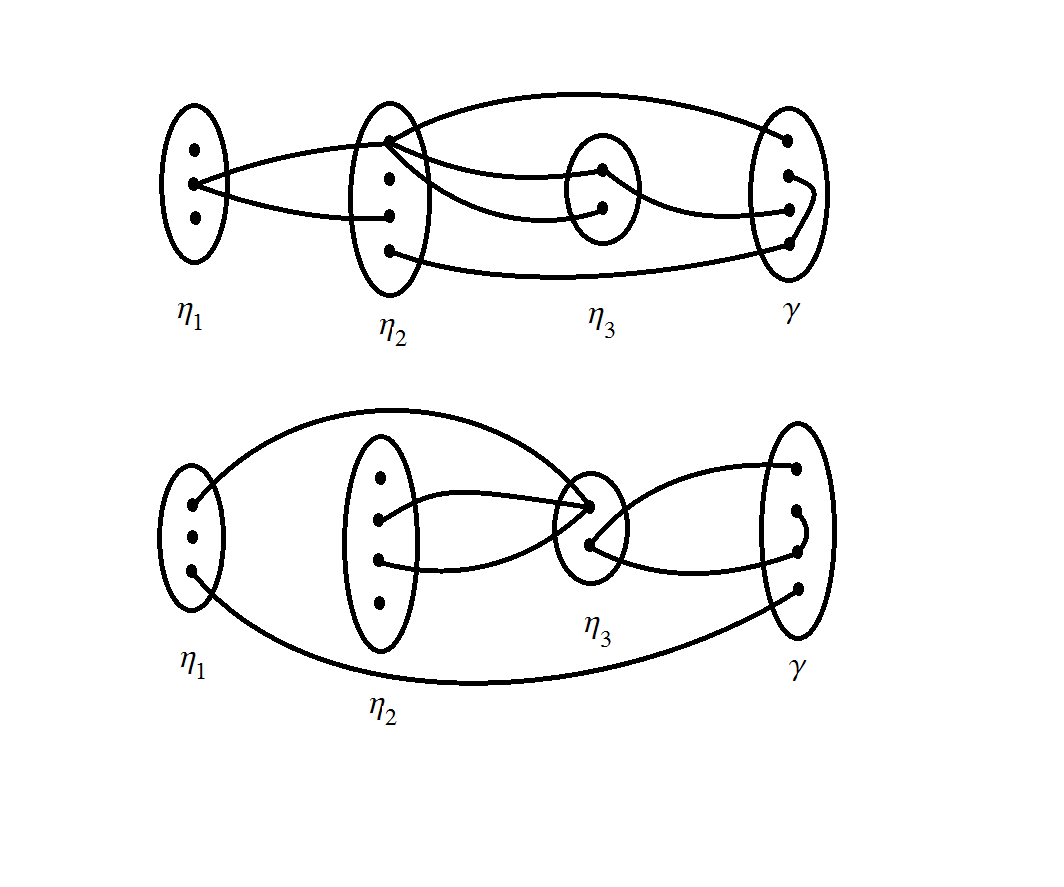}
\caption{Examples of forest graph.}
\label{fig2}
\end{figure}

\indent With the above notation, we now establish the following lemma

\begin{lemma}
\label{fitree}
The solution of the equation \eqref{Qkernel} with conditions \eqref{Qcond1}--\eqref{nukern} can be written as
\begin{equation}
\label{QGf}
Q_m(\eta_1;\dots;\eta_m\mid\gamma)=\sum_{\w{f}\in \mathfrak{S}(\eta_1;\dots;\eta_m\mid\gamma)} G_\nu(\w{f}),
\end{equation}
where the analytic contribution of a forest graph $\w{f} \in \mathfrak{S}(\eta_1;\dots;\eta_m\mid\gamma)$, denoted by $G_\nu(\w{f})$, is given by
\begin{equation}
\label{Gf}
G_\nu(\w{f})=G_\nu(\w{f};\eta_1;\dots;\eta_m\mid\gamma)=h^{l+ \vert \gamma \vert}\prod_{(x,y)\in E(\w{f})}\nu(x-y),
\end{equation}
where $E(\w{f})$ denotes the set of the edges of $\w{f}$, and where,
\begin{equation}
\label{lm}
l:=\sum_{i=1}^{m} l_{i}\,\,\,\textrm{with $l_i := \vert \eta_i\vert,\quad i=1,\dots,m$}.
\end{equation}
\end{lemma}

Individual analytic contributions are easily estimated

\begin{lemma}
\label{indivc}
Set
\begin{gather}
\label{numest0}
\nu_{0} := \max_{x\in\R^d}\nu(x) < +\infty, \\
\label{numest1}
\nu_{1} := \int_{\mathbb{R}^{d}} \nu(x)\, dx < +\infty.
\end{gather}
Then, given a forest graph $\w{f} \in \mathfrak{S}(\eta_1;\dots;\eta_m\mid\gamma)$ with $\vert \gamma \vert = n \in \mathbb{N}$,
\begin{equation}
\label{Gest}
\int_{\R^{dn}} G_\nu(\w{f};\eta_1;\dots;\eta_m\mid\{y_1,\dots, y_n\})\, dy_{1}\cdots dy_{n} \leq h^{l+ n}\nu_{0}^{\vert E_{\overline{\eta}}(\w{f})\vert}\nu_{1}^{n},
\end{equation}
where $l$ is defined in \eqref{lm}, and
\begin{equation*}
\vert E_{\overline{\eta}}(\w{f})\vert\leq l-l_1,\quad \overline{\eta}:=\bigcup_{i=1}^{m} \eta_{i},
\end{equation*}
stands for the number of edges in which one or two ends belong to the set $\bigcup_{i=1}^{m} \eta_{i}$.
\end{lemma}

It remains to estimate the number of forest graphs at fixed configurations $\bigcup_{i=1}^{m} \eta_i \cup\gamma$. We denote this number by $N_n^{(m)}(l_1,\dots, l_m)$, $l_{i}:=\vert \eta_{i}\vert$ and prove the following combinatoric lemma.

\begin{lemma}
\label{comblemma}
Let $n \in \mathbb{N}_{0}$ and $m \in \mathbb{N}$, $m\geq 2$. Set $L_i := 2^{l_i} - 1$ for $i=2,\dots,m$. Then,
\begin{equation}
\label{Numbersets}
N_n^{(m)}(l_1;\dots;l_m)= l_1(\prod_{i=2}^{m} L_i)\left(\sum_{i=1}^{m} l_{i} + n\right)^{m+n-2}.
\end{equation}
\end{lemma}

\begin{remark}
We point out that \eqref{Numbersets} is a generalization of the (well-known) Cayley formula for the number of tree graphs with $n \geq 2$ vertices
\begin{equation*}
K_n=n^{n-2},
\end{equation*}
for the case of forest graphs of a system of $m$ clusters $(\eta_i)_{i=1}^{m}$ and $n$ single vertices with  $l_i=\vert\eta_i\vert$, $n=\vert \gamma\vert$.
\end{remark}

We finally turn to\\
\noindent \textbf{Proof of Theorem \ref{exist}.} Assume that the interaction potential $\phi$ satisfies \eqref{bihzero}--\eqref{bihinfty}. Let $\beta>0$ and $z>0$ satisfying \eqref{rad}. In view of \eqref{cond3} and \eqref{numest0}--\eqref{numest1}, set
\begin{equation*}
\nu_{0}(\beta):= \max_{x \in \mathbb{R}^{d}} \nu_{\beta}(x) = \max_{x \in \mathbb{R}^{d}} \vert \mathrm{e}^{-\beta \phi(\vert x \vert)} - 1\vert < +\infty
\end{equation*}
and
\begin{equation*}
\w\nu_{0}(\beta) :=
\left\{ \begin{array}{ll} 1,  &\textrm{if $\nu_{0}(\beta)\leq 1$}, \\
\nu_{0}(\beta),  &\textrm{if $\nu_{0}(\beta)> 1$}.
\end{array}\right.
\end{equation*}
Note that $\nu_{1}(\beta)>0$ is defined in \eqref{reg}. From \eqref{TQ} along with  \eqref{QGf}, we have,
\begin{equation*}
\vert T_m(\eta_1;\dots;\eta_m \mid \ga) \vert \leq Q_m(\eta_1;\dots;\eta_m \mid \ga) =
\sum_{\w{f}\in \mathfrak{S}(\eta_1;\dots;\eta_m\mid\gamma)} G_\nu(\w{f}),
\end{equation*}
where $G_\nu(\w{f})$ is a monomial in $z$ of order $l+n$. Inserting now the bounds \eqref{Gest} and \eqref{Numbersets}, we get,
\begin{equation}
\label{easet}
\vert \w\rho^T_{m}(\eta_1;\dots;\eta_m)\vert \leq \left(2z \mathrm{e}^{2\beta B}\right)^l \left(\w\nu_{0}(\beta)\right)^{l-l_1}\sum_{n=0}^\infty\frac{(l+n)^{m+n-2}}{n!}\left(z\mathrm{e}^{2\beta B}\nu_1(\beta)\right)^n.
\end{equation}
Applying Stirling formula, i.e.,  $n!>n^n\mathrm{e}^{-n}\sqrt{2\pi n}$, we have,
\begin{equation}
\label{upbdKS}
\vert \w\rho^T_{m}(\eta_1;\dots;\eta_m)\vert \leq  \left(2z\mathrm{e}^{2\beta B+1}\right)^l \left(\w\nu_0(\beta)\right)^{l-l_1}l^{m-2} \sum_{n=0}^\infty \left(z\mathrm{e}^{2\beta B+2}\nu_1(\beta)\right)^n.
\end{equation}
Note that, we also used the bound $(1 + \frac{n}{l})^{m-2} \leq \mathrm{e}^{n}$ which follows from $l \geq m-2$.\hfill  $\blacksquare$ \\

\begin{remark}
\label{tocom}
If the condition \eqref{rad} in Theorem \ref{exist} is replaced by the condition $z \mathrm{e}^{2\beta B + 1} \nu_{1}(\beta) < 1$ as in the standard Kirkwood-Salsburg equation, a more complicated estimate for the j-PTCF can be derived as follows. From the estimate in \eqref{easet} and applying Stirling formula, we have
\begin{equation*}
\vert \w\rho^T_{m}(\eta_1;\dots;\eta_m)\vert \leq \left(2z \mathrm{e}^{2\beta B+1}\right)^l \left(\w\nu_{0}(\beta)\right)^{l-l_1}\sum_{n=0}^\infty (l+n)^{m-2}\left(z\mathrm{e}^{2\beta B+1}\nu_1(\beta)\right)^n.
\end{equation*}
The latter series obviously converges. It can in fact be estimated using the following estimate, valid for $0 \leq x <1$ and $u,v \in \mathbb{N}_0$:
\begin{equation*}
\sum_{r=v}^{\infty} r^{u} x^{r-v} \leq\sum_{k=0}^{u} \frac{k!(v+k)^{u}}{(1-x)^{k+1}}.
\end{equation*}
\end{remark}

We conclude this section by
\begin{remark}
\label{grptree}
Analogous to \eqref{QGf}, there is an analytic expression for the kernels $T_m(\eta_1;\dots;\eta_m\mid\gamma)$ in terms of forest graphs
\begin{equation}
\label{TGf}
T_m(\eta_1;\dots;\eta_m\mid\gamma) = \sum_{\w{f}\in \mathfrak{S}(\eta_1;\dots;\eta_m\mid\gamma)} G_C(\w{f}),
\end{equation}
where the contribution $C_{xy}$ for an edge of $G_C(\w{f})$ connecting vertices $x$ and $y$ is given by \eqref{Cxy}, and where the analytic expression for $G_C(\w{f})$ has the more complicated form
\begin{equation}
\label{GCf}
G_C(\w{f}) = G_C(\w{f};\eta_1;\dots;\eta_m\mid\gamma) = z^{l+n} \prod_{(x,y)\in E(\w{f})} C_{xy} \prod_{(x,y)\in (E(f^{*}) \setminus E(\w{f}))} \mathrm{e}^{-\beta\phi(\vert x-y \vert)},
\end{equation}
where $f^{*}$ denotes the maximal forest graph constructed from $\w{f}$ by adding edges connecting all points in every cluster $\eta_i$, $i=1,\dots,m$ and then adding edges using Penrose's procedure in \cite{Pe67}.
\end{remark}

\begin{remark}
Obviously, ordinary truncated (connected) correlation functions are a special case obtained by taking $\eta_1=\{x_1\},\dots,\eta_m=\{x_m\}$ in \eqref{TGf} and \eqref{GCf}. In this case, each term of the expansion is the sum of the contributions of the connected Cayley tree-graphs, and the expansion itself coincides with that obtained by O. Penrose in \cite{Pe67} (see also \cite[Eq.(4.4)]{FP07}).
\end{remark}

\subsection{Proof of Lemmas \ref{fitree}, \ref{indivc} and \ref{comblemma}.}
\label{prroof}

\noindent \textbf{Proof of Lemma \ref{fitree}.} The proof is done by induction on $n = |\gamma|$. Consider first the case $n=0$, so $\ga = \emptyset$. The equation \eqref{Qkernel} reduces then to
\begin{equation}
\label{Qmempty}
Q_m(\eta_1;\dots;\eta_m\mid\emptyset) = h \sum_{I\subset\{2,\dots, m\}} \sum^{*}_{\eta \subseteq \ov\eta_I} K_\nu(x_1;\eta) Q_{m-|I|}(\eta_1' \cup \ov\eta_I; \eta_{\{2,\dots,m\}\setminus I} \mid \emptyset).
\end{equation}
In particular, $Q_1(\eta_1 \mid \emptyset) = h Q_1(\eta_1' \mid \emptyset)$, so that $Q_1(\eta_1 \mid \emptyset) = h^{|\eta_1|}$. This agrees with \eqref{QGf} since the only allowed tree consists of individual points $x \in \eta_1$. We now do induction on $m$ and $l_1 = |\eta_1|$. For $m=1$, we already have that $Q_1(\eta_1 \mid \emptyset) = h^{l_1}$. Assuming that $Q_1 \dots,Q_{m-1}$ are given by the sum of forest contributions when $\ga =\emptyset$, the terms in \eqref{Qmempty} correspond to the construction of a forest on $\bigcup_{i=1}^{m} \eta_i$ as follows. The point $x_1$ is connected to a set of points $\eta$ outside $\eta_1$.
If $I$ is the set of indices such that $\eta \cap \eta_i \neq \emptyset$, then in $Q_{m-|I|}(\eta_1' \cup \ov{\eta}_I; \eta_{I^c} \mid \emptyset)$ there are no more connections within $\eta_1' \cup \ov{\eta}_I$, i.e., between any other points of $\eta_1$ and points of $\overline{\eta}_I$ or between two points of $\overline{\eta}_I$. In $Q_{m-|I|}$ either $m-|I| < m$ or $I = \emptyset$, in which case the first subset is $\eta'_1$ and $|\eta'_1| < |\eta_1|$. Therefore, by the induction hypothesis, its contributions are forest graphs with vertices in $\eta'_1 \cup \bigcup_{i=2}^{m} \eta_i$ such that each tree contains at most one point of $\eta'_1 \cup \ov\eta_I$. This means that when the connections with $x_1$ are added, the resulting graph still consists of separate trees.
Denote the resulting forest graph on $\bigcup_{i=1}^{m} \eta_i$ by $\w{f}$. If $x \neq x_1$ is a vertex in $\w{f}$, then by the induction hypothesis, there is a sequence of points $z_0 \in \eta'_1 \cup \bigcup_{i=2}^m \eta_i$, $z_1, \dots, z_k \in \ov\eta_{I^c}$ such that $z_k = x$ and if $z_p \in \eta_{i_p}$ ($p=0,\dots,k$) then $z_{p}$ is the unique point in $\eta_{i_p}$ connected to a point in $\eta_{i_{p+1}}$ by a edge in $\w{f}$
(note that $z_0 = x$ if $x \in \eta'_1 \cup \ov\eta_I$.) Now, if $z_0 \in \eta'_1$ or $z_0 \in \eta_i \setminus \eta$, then it is the root of a tree in $\w{f}$. If $z_0 \in \eta_i \cup \eta$ then $x_1$ is the root of the tree containing $x$ and there is no other point $x' \in \eta'_1$ connected to a point in $\eta_i$ by a edge in $\w{f}$. Collapsing the points of $\{x_1\} \cup \eta$ to a single point, the forest reduces to
a forest on $\eta'_1 \cup \ov{\eta}_I$ and $\bigcup_{i\in \{2,\dots,m\} \setminus I} \eta_i$ because there are no more edges in $\w{f}$ between points of $\eta_1 \cup \ov{\eta}_I$. The resulting forest is precisely one of the contributions to $Q_{m-|I|}(\eta'_1 \cup \ov{\eta}_I; \eta_{I^c} \mid \emptyset)$. If each $\eta_i$ is reduced to a point, the resulting graph is connected by induction except possibly in the case that $\eta_1 = \{x_1\}$ and $\eta = \emptyset$. But in that case, if $m > 1$, the contribution $Q_m(\eta'_1;\eta_2;\dots;\eta_m|\emptyset) = 0$ since $\eta'_1 = \emptyset$. The powers of $h$ are obviously correct.\\
It remains to do induction on $n$. The term $\xi = \emptyset$ gives the contribution
\begin{equation*}
h \sum_{ I\subset\{2,\dots, m\}} \sum^{*}_{\eta \subseteq \ov\eta_I} K_\nu(x_1;\eta)
Q_{m-|I|}(\eta_1' \cup \ov\eta_I; \eta_{\{2,\dots,m-k\}\setminus I} \mid \ga).
\end{equation*}
This is similar to the case $\ga = \emptyset$. It corresponds to the case where $x_1$ is only connected to points in $\bigcup_{i=2}^{m} \eta_i$, and the remaining tree after collapsing the points $\{x_1\} \cup \ov\eta_I$ gives the stated contribution by induction, since either $m-|I| < m$ or $|\eta_1'| < |\eta_1|$.
Once again the contribution of $I=\emptyset$ is zero if $\eta_1 = \{x_1\}$. The other terms are more complicated. Now, $x_1$ is connected to a set of points $\xi \subset \ga$ as well as a set of points $\eta \subset \bigcup_{i=2}^m \eta_i$. Collecting the points of $\{x_1\} \cup \xi \cup \eta$ into a single vertex, the corresponding forest is just the contribution to $Q_{m-|I|}(\eta_1' \cup \ov{\eta}_I \cup \xi; \eta_{I^c} \mid \ga \setminus \xi)$ by induction since $|\ga\setminus \xi| < n$. For $y \in \xi$, there are no more edges between other points of $\eta_1$ and $y$. Also, there are no more edges connecting $x \in \ov\eta_I$
to another point of $\eta_1$. It remains to show that upon collapsing the points of each $\eta_i$ to a single vertex, the resulting graph is a connected tree. This is more intricate. We first prove connectedness. Suppose there is an index set $J \subset \{1,\dots,m\}$ (with $1 \in J$) and a subset $\zeta \subset \ga$ such that $J \neq \{1,\dots,m\}$ or $\zeta \neq \ga$ and there are no edges between points of $\overline{\eta}_J \cup \zeta$ and points in the complement. Clearly, $I \subset J$ since the points of $\eta$ are connected to
$x_1$ (due to the factor $K_\nu(x_1;\xi \cup \eta)$) and $\eta \cap \eta_i \neq \emptyset$ for $i \in I$. Also $\xi \subset \zeta$ for the same reason. By the induction hypothesis for $Q_{m-|I|}(\eta'_1 \cup \ov\eta_I \cup \xi; \eta_{I^c} \mid \ga \setminus \xi)$, there are no sets $J' \supset I$ and $\zeta' \subset \ga\setminus \xi$ without external edges, other than the trivial $J'=\emptyset$ and $\zeta' = \emptyset$ or
$J' = I \cup I^c$ and $\zeta'=\ga \setminus \xi$. Therefore $J \supset \{2,\dots,m\}$ and $\ga \setminus \xi \subset \zeta$, or $J \cap \{2,\dots,m\} = \emptyset$ and $\zeta \subset \ga\setminus \xi$. In the first case $J = \{1,\dots,m\}$ and $\zeta = \ga$, which contradicts the initial assumption. In the second case, $J = \{1\}$ and $\xi = \emptyset$, and since $I \subset J$, $I = \emptyset$. The corresponding contribution
equals zero as above, since then $\eta'_1 \cup \ov\eta_I \cup \xi = \emptyset$. To see that the resulting graph is a tree, note that in any contributing forest to $Q_{m-|I|}(\eta'_1 \cup \xi \cup \ov\eta_I; \eta_{I^c} \mid \ga\setminus \xi)$ there is just one edge between a point of $\eta'_1 \cup \xi \cup \ov\eta_I$ and a tree on $\eta_{I^c} \cup \ga \setminus \xi$. The factor $K_\nu(x_1;\xi \cup \eta)$ gives edges between $x_1$ and the points of $\xi \cup \eta$, and therefore to only one point of this tree. \hfill  $\blacksquare$ \\

\noindent \textbf{Proof of Lemma \ref{indivc}.} We only give the main arguments. If $y_i$ is an end vertex of a tree in the forest $\w{f}$, then a contribution involving the factor $\nu_1$ arises. The same holds if, from $y_i$ outwards, there are only vertices $y_k$ since we can integrate them successively. In the case when $y_i$ lies between the points $x_i$ and $x_k$ in $\bigcup_{i=1}^{m} \eta_i$, we first use the inequality $\nu(y_i - x_k)\leq\nu_0$.  \hfill $\blacksquare$ \\

\noindent \textbf{Proof of Lemma \ref{comblemma}.} From \eqref{Qkernel}, it can be seen that $N_n^{(m)}(l_1;\dots;l_m)$ satisfies the recurrent relations
\begin{equation}
\label{Numbersetst1}
N_n^{(m)}(l_1;\dots;l_m)=\sum_{k=0}^n {n \choose k} \sum_{I\subset\{2,\dots,m\}} L_I N_{n-k}^{(m-|I|)}(l_1+l_I+k-1;l_{i_2};\dots;l_{i_{m-|I|}}),
\end{equation}
where we denote $L_I:=\prod_{i\in I}L_i$, $l_I:=\sum_{i\in I}l_i$ (with the convention $l_{\emptyset} := 0$) and $\{i_2,\dots,i_{m-|I|}\} := \{2,\dots,m\} \setminus I$. Hereafter, we set $l := \sum_{i=1}^{m} l_i$. Let us introduce new numbers $\tilde{N}_n^{(m)}(l_1;\dots;l_m)$ in such a way that
\begin{equation*}
N_n^{(m)}(l_1;\dots;l_m)= (\prod_{i=2}^{m} L_{i}) \tilde{N}_n^{(m)}(l_1;\dots;l_m).
\end{equation*}
Then, the recurrent relations \eqref{Numbersetst1} can be rewritten in the following way,
\begin{equation}
\label{Numbersetst3}
\tilde{N}_n^{(m)}(l_1;\dots;l_m)=\sum_{k=0}^n {n \choose k} \sum_{I\subset\{2,\dots,m\}}
\tilde{N}_{n-k}^{(m-|I|)}(l_1+l_I+k-1;l_{i_2};\dots;l_{i_{m-|I|}}).
\end{equation}
We now prove that, given the initial condition $\tilde{N}_0^{(1)}(l_1) = 1$,
\begin{equation}
\label{solution}
\tilde{N}_n^{(m)}(l_1;\dots;l_m)=l_1(\sum_{i=1}^{m} l_{i} + n)^{m+n-2}
\end{equation}
is the solution of the recurrent relations \eqref{Numbersetst3}. Inserting the identity
\begin{equation*}
\tilde{N}_{n-k}^{(m-|I|)}(l_1+l_I+k-1;l_{i_2};\ldots;l_{i_{m-|I|}})=(l_1+l_I+k-1)(l+n-1)^{m+n-k-|I|-2},
\end{equation*}
in the right-hand side of \eqref{Numbersetst3}, we obtain,
\begin{equation*}
\tilde{N}_n^{(m)}(l_1;\dots;l_m) = \sum_{i=1}^{3} M_{i},
\end{equation*}
where,
\begin{align*}
M_1 :={}& \sum_{k=0}^n {n \choose k}(l+n-1)^{n-k-1}\sum_{I\subset\{2,\dots, m\}} l_1 (l+n-1)^{m-|I|-1},\\
M_2 :={}&  \sum_{k=0}^n {n \choose k}(l+n-1)^{n-k-1}\sum_{I\subset\{2,\dots, m\}} l_I (l+n-1)^{m-|I|-1}, \\
M_3 :={}&  \sum_{k=0}^n {n \choose k}(l+n-1)^{n-k-1}\sum_{I\subset\{2,\dots, m\}} (k-1) (l+n-1)^{m-|I|-1}.
\end{align*}
Now, in $M_2$ we first sum over sets $I$ with $|I|=p$ using
\begin{equation*}
\sum_{\substack{I \subset \{2,\dots,m\} \\ |I|=p}} l_I = {m-2 \choose p-1} \sum_{i=2}^m l_i = {m-2 \choose p-1} (l-l_1).
\end{equation*}
In the other two sums, this summation is easy, and we obtain
\begin{align*}
M_1 &= l_1 \sum_{k=0}^n{n \choose k}(l+n-1)^{n-k-1}\sum_{p=0}^{m-1} {m-1 \choose p}  (l+n-1)^{m-p-1} \\
&= l_1 \sum_{k=0}^n {n \choose k} (l+n-1)^{n-k-1} (l+n)^{m-1} = l_1 (l+n-1)^{-1} (l+n)^{m+n-1},\\
M_2 &= (l-l_1) \sum_{k=0}^n{n \choose k}(l+n-1)^{n-k-1}\sum_{p=1}^{m-1} {m-2 \choose p-1}  (l+n-1)^{m-p-1} \\
&= (l-l_1) (l+n-1)^{-1} (l+n)^{m+n-2},\\
M_3 &= \sum_{k=0}^n{n \choose k}(l+n-1)^{n-k-1}\sum_{p=0}^{m-1} {m-1 \choose p} (k-1) (l+n-1)^{m-p-1} \\
&= \sum_{k=0}^n {n \choose k} (k-1) (l+n-1)^{n-k-1} (l+n)^{m-1} = -l (l+n-1)^{-1} (l+n)^{m+n-2},
\end{align*}
where we used the identity
\begin{equation*}
\sum_{k=0}^n{n \choose k}(k-1)(l+n-1)^{n-k-1}\;=\;-l(l+n-1)^{-1}(l+n)^{n-1}.
\end{equation*}
We conclude that $\sum_{i=1}^{3} M_i = l_1(l+n)^{m+n-2}$ which completes the induction. This proves \eqref{solution}. \hfill $\blacksquare$

\section{Strong decay properties for PTCF.}
\label{strdca}
\setcounter{equation}{0} \renewcommand{\theequation}{\arabic{section}.%
\arabic{equation}} 

Theorem \ref{exist} states the existence of a unique solution to the equation \eqref{FunctPTCmnew} in the form of convergent expansions, see \eqref{rhomj=1} with Remark \ref{grptree} and also \eqref{TQ} with \eqref{QGf}--\eqref{Gf}. The most important property of the PTCF is their decay as the distances between the clusters increases, i.e., $\dist(\eta_i, \eta_j)\rightarrow +\infty$, $i\neq j$.

\subsection{Polynomial decay for PTCF.}

We start by formulating the main result of Sec. \ref{strdca}

\begin{theorem}
\label{polynom}
Suppose that the interaction potential $\phi$ satisfies \eqref{stU} and \eqref{reg}. Assume in addition that there exists $\alpha>d$ and, for all $\beta>0$, there exists a constant $C(\beta)>0$ such that
\begin{equation*}
\nu_{\beta}(x) := \vert \mathrm{e}^{-\beta \phi(x)}-1 \vert \leq C(\beta) \overline{\nu}(x),
\end{equation*}
with
\begin{equation}
\label{polykernel}
\overline{\nu}(x):= \frac{1}{1 + \vert x\vert^{\alpha}},\quad x \in \mathbb{R}^{d}.
\end{equation}
Then, provided that,
\begin{equation*}
z \mathrm{e}^{2\beta B} [\nu_1(\beta) \mathrm{e} + \overline{\nu}_{1}(\beta) (\mathrm{e} + 2^{1+\alpha})] < 1,
\end{equation*}
where $B \geq 0$ is defined in \eqref{stU}, $\nu_{1}(\beta)>0$ in \eqref{reg} and
\begin{equation*}
\overline{\nu}_{1}(\beta) := C(\beta)\overline{\nu}_{1},\quad \overline{\nu}_{1} := \int_{\mathbb{R}^{d}} \overline{\nu}(x)\,dx < +\infty,
\end{equation*}
there exist, given $m\in\mathbb{N}$ with $m\geq 2$, constants $A_{m,\sigma}=A_{m,\sigma}(\beta,z,\alpha) > 0$, $1 \leq \sigma \leq m$ such that the PTCF in \eqref{rhomj=1} admit the following bounds
\begin{equation*}
\vert \tilde{\rho}^T_m(\eta_1; \dots; \eta_m)\vert \leq \sum_{\sigma=1}^{m} A_{m,\sigma} \max_{T_{m} \in \mathcal{T}_{m}} \overline{\nu}_{T_{m}},
\end{equation*}
where $\mathcal{T}_{m}$ denotes the set of trees on $m$ points, and
\begin{equation*}
\overline{\nu}_{T_{m}} := \prod_{(i,j) \in T_{m}} \max_{x_{i} \in \eta_{i}; x_{j} \in \eta_{j}} \overline{\nu}(x_{i} - x_{j}).
\end{equation*}
\end{theorem}

\begin{remark}
Explicit upper bounds for the constants $A_{m,\sigma}$ are derived in the proof. Setting \begin{equation*}
z \mathrm{e}^{2\beta B}=h,\quad \nu_{1}(\beta)=\nu_{1},\quad C(\beta)=C,
\end{equation*}
$A_{2,\sigma}$ with $\sigma =1,2$ are given in \eqref{A12}, $A_{3,\sigma}$ with $\sigma=1,2,3$ are given in \eqref{A31}, \eqref{A32} and \eqref{A33}, and for any $m\geq 4$, $A_{m,1}$, $A_{m,2}$ and $A_{m,\sigma}$ with $3\leq \sigma \leq m$ are given in \eqref{Am1}, \eqref{Am2} and \eqref{Amm} respectively. To derive these upper bounds,  we use the combinatoric identities \eqref{combidt}.
\end{remark}

The rest of Sec. \ref{strdca} is devoted to the proof of Theorem \ref{polynom}. It is organized as follows. We first establish two technical results, see Lemma \ref{decaylemma} and Proposition \ref{estQ_m} below. Subsequently, we prove Theorem \ref{polynom} in the case $m=2$, $m=3$ and the general case $m\geq 4$ in Sec. \ref{m=2}, \ref{m=3} and \ref{m=gen} respectively.\\
\indent We point out that, from \eqref{rhomj=1} with \eqref{Tkernel} (and the conditions \eqref{cond1}--\eqref{cond2}), it is sufficient, by virtue of Lemma \ref{riglem}, to work with the family of kernels in \eqref{Qkernel} (with the conditions \eqref{Qcond1}--\eqref{nukern}). \\
\indent Consider for instance forest graphs $\w{f} \in \mathfrak{S}(\eta_1;\dots;\eta_m\mid\{y_{1}\})$. Restricting the diagram to $\bigcup_{i=1}^{m} \eta_i$, one obtains a forest on $\bigcup_{i=1}^{m} \eta_i$ of which some trees are connected by an edge in $\w{f}$ to $y_1$. If there is just one such edge, the corresponding contribution is obtained from that of the restricted forest by multiplying by $\nu(x_j-y_1)$ if $x_j$ is the vertex attached to $y_1$. In general, one has to multiply by a factor $\prod_{r=1}^p \nu(x_{j_r} - y_1)$. In the former case, integration with respect to the variable $y_1$ simply multiplies the contribution of the diagram from $\mathfrak{S}(\eta_1;\dots;\eta_m\mid\emptyset)$ by the factor $\nu_{1}$, see \eqref{numest1}. In the general case, we need to consider integrals of the form
\begin{equation}
\label{2estm}
\int_{\mathbb{R}^{d}} \prod_{r=1}^p \nu(x_{r} - y)\,dy.
\end{equation}
In case that the kernel $\nu$ has a polynomial decay, terms of type \eqref{2estm} can be easily estimated

\begin{lemma}
\label{decaylemma}
Let $\overline{\nu}$ be the kernel in \eqref{polykernel} with $\alpha>d$. Then, for all $p \in \mathbb{N}$, $p \geq 2$ and $x_1, \dots, x_p \in \R^d$,
\begin{equation}
\label{nuineq}
\int_{\mathbb{R}^{d}} \prod_{r=1}^{p} \ov{\nu}(x_r - y)\,dy \leq 2^{\alpha(p-1)} \overline{\nu}_{1} \sum_{r=1}^p \prod_{\substack{k=1 \\ k\neq r}}^{p} \ov{\nu}(x_k-x_r).
\end{equation}
\end{lemma}

\noindent \textbf{Proof.} We subdivide the integral with respect to $y$ into domains where $|y-x_r| <  \max_{k\neq r} |y-x_k|$. Then, $|x_k - y| > \frac{1}{2} |x_k-x_r|$ and the inequality \eqref{nuineq} easily follows from
\begin{equation*}
\frac{1}{\vert x_k-y \vert^\alpha + 1} < \frac{1}{(\frac{1}{2} \vert x_k-x_r \vert)^\alpha + 1} < \frac{2^\alpha}{\vert x_k-x_r \vert^\alpha + 1}. \tag*{\hfill $\blacksquare$}
\end{equation*}

To count the possible diagrams, we will first isolate the parts of the diagram consisting of trees with vertices in $\ga$ except possibly one endpoint. This can be done as follows. Define
\begin{equation}
\label{Qcond}
\begin{split}
Q_m(\eta_1;\dots;\eta_m\mid 0) :={}& Q_m(\eta_1;\dots;\eta_m\mid \emptyset),\\
Q_m(\eta_1;\dots;\eta_m\mid n) :={}& \int_{\mathbb{R}^{dn}}  Q_m(\eta_1;\dots;\eta_m \mid \{y_1,\dots, y_n\})\,dy_{1} \cdots dy_{n}, \quad n \in \mathbb{N},
\end{split}
\end{equation}
where the family of kernels $Q_m(\eta_1;\dots;\eta_m \mid \gamma)$, $m\geq 2$ and $\gamma \in \Gamma_{0}$ is given in \eqref{Qkernel} with the conditions \eqref{Qcond1}--\eqref{nukern}. It then satisfies the following recursion relation
\begin{multline}
\label{Qkernel-2}
Q_m(\eta_1;\dots;\eta_m\mid n) = h \sum_{ I\subset\{2,\dots,m\}} K^{(0)} (x_1;\overline{\eta}_I) \\
\times \sum_{k=0}^n {n \choose k}\int_{\mathbb{R}^{d k}} \prod_{j=1}^k K_\nu(x_1;y_j)Q_{m-\vert I \vert}(\eta_{1}'\cup \overline{\eta}_I \cup \{y_1,\dots, y_k\};\eta_{\{2,\dots,m\}\setminus I} \mid n-k)\,dy_1\cdots dy_k,
\end{multline}
where it is understood that the term $k=0$ in the sum reduces to $Q_{m-\vert I \vert}(\eta_{1}'\cup \overline{\eta}_I;\eta_{\{2,\dots,m\}\setminus I} \mid n)$, and
\begin{equation}
\label{Qcond3}
\begin{split}
&K^{(0)}(x_i;\emptyset):= 1,\quad i \in \{1,\dots,m\},\\
&K^{(0)}(x_i;\overline{\eta}_I):= \sum^{*}_{\eta\subset \overline\eta_I} K_\nu(x_i;\eta)=
\sum^{*}_{\eta\subset \overline\eta_I}\prod_{x\in\eta}\nu(x_i-x),\quad I \subset \{1,\dots,m\}\setminus\{i\}.
\end{split}
\end{equation}

We then establish
\begin{prop}
\label{estQ_m}
Given $n \in \mathbb{N}_{0}$, the solution of the recursion relation \eqref{Qkernel-2} can be expressed as
\begin{equation*}
Q_m(\eta_1;\dots;\eta_m\mid n) = h^l \sum_{k=0}^{n} {n \choose k}
k! N^{(1)}_{n-k}(l+k) (h\nu_1)^{n-k} \widetilde{Q}_m(\eta_1;\dots;\eta_m \mid k),
\end{equation*}
where $l$ is defined in \eqref{lm}, $\nu_{1}$ in \eqref{numest1}, $N^{(1)}_{k'}$ with $0\leq k' \leq n$ is given by \eqref{Numbersets}, and $\widetilde{Q}_m(\eta_{1};\dots;\eta_m \mid k)$ consists of the contributions from all forest graphs in $\mathfrak{S}(\eta_1;\dots;\eta_m\mid \{y_1,\dots, y_k\})$ in which all vertices of $y_i$ of $\ga$ are connected to at least two other vertices.
\end{prop}

\noindent \textbf{Proof}. This can be proved inductively from the formula \eqref{Qkernel-2}. However, it is also easily understood graphically as follows. Given a forest graph in $\mathfrak{S}(\eta_1;\dots;\eta_m\mid \{y_1,\dots, y_n\})$, consider the points of $\ga = \{y_1,\dots,y_n\}$ connected to only one other vertex (endpoints). These are parts of trees on $\ga$
with a single base point either in $\ga$ or in $\bigcup_{i=1}^{m}\eta_i$. Starting at the endpoints, the
corresponding points $y_i$ can easily be integrated, yielding factors $h\nu_1$. In the remaining graph, each point of $\ga$ is connected to at least two other vertices. We denote the contribution of this graph by
$\widetilde{Q}_m(\eta_{1};\dots;\eta_m \mid k)$, where $k$ is the number of remaining vertices in $\ga$.
Conversely, given a forest graph in $\mathfrak{S}(\eta_1;\dots;\eta_m\mid \{y_1,\dots, y_k\})$ in which each point $y_i$ ($i=1,\dots,k$) is connected to at least two other vertices, we obtain the contribution from graphs in $\mathfrak{S}(\eta_1;\dots;\eta_m\mid \{y_1,\dots, y_n\})$ with $n \geq k$, containing this graph and such that all other points $y_{k+1}, \dots, y_n$ are in trees with a single base point, by counting the number of possibilities of attaching trees to the given tree with total number of vertices equal to $n-k$. But this number is given precisely by
\begin{equation*}
{n \choose k} k! N^{(1)}_{n-k}(l+k) (h\nu_1)^{n-k}.
\end{equation*}
Indeed, we can choose which of the total of $n$ points belongs to the original graph in $n \choose k$ ways and order them in $k!$ ways. The number of ways of forming trees out of the remaining $n-k$ points is then given by $N^{(1)}_{n-k}(l+k)$, because for this purpose we can  consider all points of the original graph as belonging to a single cluster as they cannot be connected further to each other. There are obviously $l+k$ such points to be connected to a further $n-k$ external points. By Lemma~\ref{comblemma}, this can be done in $N^{(1)}_{n-k}(l+k)$ ways. \hfill $\blacksquare$

\subsection{The case $m=2$.}
\label{m=2}

There are two possibilities: either there is at least one edge between $\eta_1$ and $\eta_2$ in the forest, or there is none. In the first case, the restriction of the forest to $\ga$ splits into separate trees, each of which is connected to a single point of either $\eta_1$ or $\eta_2$. In the second case, the restriction to $\ga$ also splits into separate trees, but one of these is connected to a single point of $\eta_1$ as well as one or more points of $\eta_2$. The others are again connected to a single point of either $\eta_1$ or $\eta_2$. The trees connected to a single point are easily integrated out, giving rise to factors $\nu_1$. If there is a tree connecting $\eta_1$ and $\eta_2$ then there is one point $y_1$ of that tree in $\ga$ connected to a point of $\eta_1$ and one point $y_2 \in \ga$ connected to one or more points of $\eta_2$ ($y_1$ can be equal to $y_2$). In that case, there is a unique path in the tree connecting $y_1$ to $y_2$. The remaining part of the tree consists of individual trees connected to single points of this path (or points of $\eta_1 \cup \eta_2$). These can be integrated out giving factors $\nu_1$ as before. In terms of Proposition \ref{estQ_m},
\begin{equation}
\label{Q2n}
Q_2(\eta_1;\eta_2\mid n) = h^l \sum_{k=0}^n {n \choose k}
k! N^{(1)}_{n-k}(l+k) (h\nu_1)^{n-k} \widetilde{Q}_2(\eta_1;\eta_2 \mid k),\quad n \in \mathbb{N}_{0},
\end{equation}
with
\begin{equation*}
\widetilde{Q}_2(\eta_1;\eta_2 \mid k) := \sum_{x_1 \in \eta_1} K^{(k)}(x_1;\eta_2),\quad 0 \leq k \leq n,
\end{equation*}
where $K^{(0)}(x_i;\eta_j)$, $x_{i} \in \eta_{i}$ and $i\neq j$ is given in \eqref{Qcond3} and $K^{(k)}(x_i;\eta_j)$, $x_{i} \in \eta_{i}$ and $i\neq j$ are defined as
\begin{equation}
\label{estK2}
K^{(k)}(x_i;\eta_j) := h^k \int_{\mathbb{R}^{dk}} \nu(x_i-y_1) \prod_{r=1}^{k-1}
\nu(y_r-y_{r+1}) K^{(0)}(y_k;\eta_j)\,dy_1 \cdots dy_k,\quad k \geq 1.
\end{equation}
Assume now that $\nu$ is polynomially bounded, i.e. $\nu(x) \leq C \ov{\nu}(x)$ with $\ov{\nu}$ in \eqref{polykernel} for some constant $C > 0$ and $\alpha >d$. Integrating over the points on the path from $y_1$ to $y_{k-1}$, Lemma \ref{decaylemma} yields factors $2^{1+\alpha} C \ov{\nu}_1$
\begin{equation}
\label{Kkbd}
K^{(k)}(x_1;\eta_2) \leq (hC)^k (2^{1+\alpha} \ov{\nu}_1)^{k-1} \int_{\mathbb{R}^{d}} \ov{\nu}(x_1-y) K^{(0)}(y;\eta_{2})\, dy,\quad k \geq 1.
\end{equation}
Here, we also used the bound $\overline{\nu} \leq 1$. The integral in \eqref{Kkbd} can be estimated as follows. From \eqref{Qcond3},
\begin{equation}
\label{fansum0}
K^{(0)}(y;\eta_{i}) \leq \sum_{x_{i} \in \eta_{i}} \sum_{\eta' \subset (\eta_{i} \setminus \{x_{i}\})} C^{\vert \eta' \vert + 1} \ov{\nu}(y-x_{i}) \leq C(1+C)^{l_{i}-1} \sum_{x_{i} \in \eta_{i}} \ov{\nu}(y-x_{i}).
\end{equation}
Inserting \eqref{fansum0} in \eqref{Kkbd} and then using Lemma \ref{decaylemma} again, we obtain the common upper bound
\begin{equation}
\label{Kkineq}
K^{(k)}(x_1;\eta_2) \leq  (h \ov{\nu}_1 2^{1+\alpha} C)^k C(1+C)^{l_2-1} \sum_{x_2 \in \eta_2} \ov{\nu}(x_1-x_2),\quad 0\leq k \leq n.
\end{equation}
In summing over the trees connected to a single point of this path, the number of vertices in these trees is unlimited. This means that we can consider these trees individually, having base points on the $k+l$ points of the path from $\eta_1$ to $\eta_2$ and containing $n_i + 1$ points ($i=1,\dots,k+l$). There are $(n_i+1)^{n_i-1}$ such trees for each $i$, so we now have in total,
\begin{equation*}
\vert \w\rho^T_{2}(\eta_1;\eta_2)\vert
\leq   C (1+C)^{l_2-1} h^l \sum_{x_1 \in \eta_1} \sum_{k=0}^\infty
(h \ov{\nu}_1 2^{1+\alpha} C)^k \sum_{n_1,\dots,n_{k+l}=0}^\infty \prod_{i=1}^{k+l} \frac{(n_i+1)^{n_i-1} (h \nu_1)^{n_i}}{n_i!} \sum_{x_2 \in \eta_2} \ov{\nu}(x_1-x_2).
\end{equation*}
Here, there is a factor $\frac{n!}{k! n_1! \dots n_{k+l}!}$ for the number of ways of distributing the vertices in $\ga$ over the individual trees and the remaining $k$ points of $\ga$ and a factor $k!$ for the number of ways of ordering the vertices in the path connecting the two clusters as well as a factor $\frac{1}{n!}$ from the definition of the correlation function. Using now that $(k+1)^{k-1} \leq  k! \mathrm{e}^k$ for $k\geq 1$, we obtain,
\begin{equation*}
\vert \w\rho^T_{2}(\eta_1;\eta_2)\vert \leq l_1 l_2  C (1+C)^{l_2-1}  h^l \sum_{k=0}^\infty \left(h \ov{\nu}_1 2^{1+\alpha} C\right)^k \left(\sum_{n=0}^\infty (h \nu_1 \mathrm{e})^n\right)^{k+l}  \max_{x_1 \in \eta_1;\,x_2 \in \eta_2} \ov{\nu}(x_1-x_2).
\end{equation*}
Here, we assumed that $ h (\nu_1 \mathrm{e} + \ov{\nu}_1 2^{1+\alpha} C) < 1$. Theorem \ref{polynom} in the case $m=2$ is proven by setting
\begin{equation}
\label{A12}
A_{2,1} = A_{2,2} := \frac{1}{2} l_1 l_2 C (1+C)^{l_2-1}  \left(\frac{h}{1-h \nu_1 \mathrm{e}}\right)^l \frac{1-h \nu_1 \mathrm{e}}{1- h \nu_1 \mathrm{e} - h \ov{\nu}_1 2^{1+\alpha} C}.
\end{equation}
\begin{remark}
Comparing the above formula with expression \eqref{Q2n}, we have the remarkable identity,
\begin{equation*}
N^{(1)}_{n}(l) = \sum_{\substack{n_1,\dots,n_{l} \geq 0 \\ \sum\limits_{i=1}^{l} n_i = n}} \frac{n!}{n_1! \cdots n_l!} \prod_{i=1}^l (n_i+1)^{n_i-1},
\end{equation*}
where we replaced $n-k$ by $n$ and $k+l$ by $l$.
\end{remark}

\subsection{The case $m=3$.}
\label{m=3}

Here, the situation is not too much more complicated. The cases where there is a edge between at least one pair of $\eta_1$, $\eta_2$ and $\eta_3$ reduce to the case $m=2$. There remains the case that there is a tree on $\ga$ which is connected to all three. Again, this tree has only one point in $\ga$ which connects to $\eta_i$ for each $i=1,2,3$, and by integrating out over intermediate $y$'s which connect to only two others, this reduces to the case where these three points coincide. Assuming that the points connecting the tree to $\eta_1$, $\eta_2$ and $\eta_3$ are different points $y_1$, $y_2$ and $y_3$, there are 3 possible permutations of these points, and we can integrate out any intermediate points as before, yielding factors $2^{1+\alpha} \ov{\nu}_1$. In terms of Proposition \ref{estQ_m}, we have,
\begin{equation}
\label{Q3n}
Q_3(\eta_1;\eta_2;\eta_3\mid n) = h^l \sum_{k=0}^n {n \choose k} k! N^{(1)}_{n-k}(l+k) (h\nu_1)^{n-k} \widetilde{Q}_3(\eta_1;\eta_2;\eta_3 \mid k),\quad n \in \mathbb{N}_{0},
\end{equation}
where $\widetilde{Q}_3(\eta_1;\eta_2;\eta_3 \mid k)$ is the contribution from all forest graphs in
$\mathfrak{S}(\eta_1;\eta_{2};\eta_3\mid \{y_1,\dots, y_k\})$, in which all vertices $y_i$ of $\ga$ are connected to at least two other vertices. Integrating out the vertices of $\ga$ connected to only 2 others yields factors $2^{1+\alpha} \ov{\nu}_1$ and results in a tree on $\ga$ where every vertex is connected to at least 3 others. There is only one such tree. It consists of a single point $y$ of $\ga$ connected to $\eta_1$, $\eta_2$ and $\eta_3$. Conversely, given this tree, one can form trees with additional vertices connected to two points by adding a sequence of points between $y$ and $\eta_1$, $\eta_2$ and $\eta_3$. In total, $\widetilde{Q}_3(\eta_1;\eta_2;\eta_3 \mid k)$ is the sum of 3 contributions
\begin{equation}
\label{sumthree}
\widetilde{Q}_3(\eta_1;\eta_2;\eta_3 \mid k) = \widetilde{Q}_{3,1}(\eta_1;\eta_2;\eta_3 \mid k) +
\widetilde{Q}_{3,2}(\eta_1;\eta_2;\eta_3 \mid k) + \widetilde{Q}_{3,3}(\eta_1;\eta_2;\eta_3 \mid k), \end{equation}
where $\widetilde{Q}_{3,3}(\eta_1;\eta_2;\eta_3 \mid k)$ contains the contributions of terms where there is no
connection inside  $\eta_1 \cup \eta_2 \cup \eta_3$ (3 components), $\widetilde{Q}_{3,2}(\eta_1;\eta_2;\eta_3 \mid k)$ corresponds to the terms where there is one or more  edge(s) between one pair of $\eta_1$, $\eta_2$ and $\eta_3$ (2 components), and $\widetilde{Q}_{3,1}(\eta_1;\eta_2;\eta_3 \mid k)$ contains the contributions where all 3 clusters are connected by edges inside $\eta_1 \cup \eta_2 \cup \eta_3$. In the latter, we must have $k=0$ since there cannot be another (outside) connection between two $\eta_i$'s. From \eqref{sumthree}, \eqref{Q3n} is the sum of three contributions
\begin{equation*}
\begin{split}
Q_{3,1}(\eta_1;\eta_2;\eta_3\mid n) :=&{} h^l \widetilde{Q}_{3,1}(\eta_1;\eta_2;\eta_3 \mid 0), \\
Q_{3,i}(\eta_1;\eta_2;\eta_3\mid n) := &{} h^l \sum_{k=0}^n {n \choose k} k! N^{(1)}_{n-k}(l+k) (h\nu_1)^{n-k}\widetilde{Q}_{3,i}(\eta_1;\eta_2;\eta_3 \mid k),\quad i=2,3.
\end{split}
\end{equation*}
\indent In the case when all three clusters are connected,
\begin{equation*}
\begin{split}
\widetilde{Q}_{3,1}(\eta_1;\eta_2;\eta_3 \mid 0) = &\sum_{x_1 \in \eta_1} \sum_{\eta'_2 \subset \eta_2}^{*} \sum_{\eta'_3 \subset \eta_3}^{*} \left\{ \prod_{x_2 \in \eta'_2} \nu(x_1-x_2) \sum_{x'_1 \in \eta_1}
\prod_{x_3 \in \eta'_3} \nu(x'_1-x_3) \right.\\ &+ \left. \prod_{x_2 \in \eta'_2} \nu(x_1-x_2) \sum_{x'_2 \in \eta_2}
\prod_{x_3 \in \eta'_3} \nu(x'_2-x_3) + \prod_{x_3 \in \eta'_3} \nu(x_1-x_3) \sum_{x'_3 \in \eta_3}
\prod_{x_2 \in \eta'_2} \nu(x'_3-x_2)\right\}.
\end{split}
\end{equation*}
Assume that $\nu(x) \leq C \ov{\nu}(x)$ with $\ov{\nu}$ as in \eqref{polykernel} for some constant $C > 0$ and $\alpha>d$. From \eqref{fansum0} together with the following upper bound on the sum over trees
\begin{equation}
\label{treesum}
h^l \sum_{n=0}^\infty \frac{(h\nu_1)^{n}}{n!} N^{(1)}_{n}(l) \leq h^l \sum_{n_1,\dots,n_l = 0}^\infty \prod_{i=1}^l \frac{(h \nu_1)^{n_i}}{n_i!} (n_i+1)^{n_i-1} \leq h^l \prod_{i=1}^l \sum_{n_i=0}^\infty (h \nu_1 \mathrm{e})^{n_i} = \left(\frac{h}{1-h \nu_1 \mathrm{e}}\right)^l,
\end{equation}
and
\begin{equation*}
\sum_{x_{i} \in \eta_{i}} \sum_{x_{j} \in \eta_{j}} \overline{\nu}(x_{i}-x_{j}) \leq l_{i} l_{j} \max_{x_i \in \eta_i;\,x_j \in \eta_j} \ov{\nu}(x_i-x_j), \quad i\neq j,
\end{equation*}
we then obtain,
\begin{align}
\sum_{n=0}^\infty &\frac{1}{n!} Q_{3,1}(\eta_1;\eta_2,\eta_3 \mid n) \nonumber\\
\leq&  l_1 l_2 l_3 C^{2} (1+C)^{l_2+l_3-2} \left(\frac{h}{1-h \nu_1 \mathrm{e}}\right)^l \left \{ l_1 \max_{x_1 \in \eta_1;\,x_2 \in \eta_2} \ov{\nu}(x_1-x_2) \max_{x_1 \in \eta_1;\,x_3 \in \eta_3} \ov{\nu}(x_1-x_3) \nonumber \right.\\
&+ \left.l_2 \max_{x_1 \in \eta_1;\,x_2 \in \eta_2} \ov{\nu}(x_1-x_2) \max_{x_2 \in \eta_2;\,x_3 \in \eta_3} \ov{\nu}(x_2-x_3) + l_3 \max_{x_1 \in \eta_1;\,x_3 \in \eta_3} \ov{\nu}(x_1-x_3)
\max_{x_3 \in \eta_3;\,x_2 \in \eta_2} \ov{\nu}(x_3-x_2)\right\} \nonumber \\
\label{rho31}
\leq& l_1 l_2 l_3 (l_{1}+l_{2}+l_{3}) C^{2} (1+C)^{l_2+l_3-2} \left(\frac{h}{1-h \nu_1 \mathrm{e}}\right)^l \max_{T_{3} \in \mathcal{T}_{3}} \overline{\nu}_{T_{3}}.
\end{align}
\indent In the cases where only one pair of $\eta_1$, $\eta_2$ and $\eta_3$ are connected, we have, for all $0\leq k \leq n$,
\begin{align*}
\widetilde{Q}_{3,2}(\eta_1;\eta_2;\eta_3 \mid k) := \sum_{x_1 \in \eta_1} &\left\{ K^{(0)}(x_1;\eta_2)
\sum_{x \in \eta_1 \cup \eta_2} K^{(k)} (x;\eta_3) + K^{(0)}(x_1;\eta_3) \sum_{x \in \eta_1 \cup \eta_3} K^{(k)} (x;\eta_2) \right.\\
&+\left. K^{(k)}(x_1;\eta_2) \sum_{x_2 \in \eta_2} K^{(0)}(x_2,\eta_3) + K^{(k)}(x_1;\eta_3) \sum_{x_3 \in \eta_3} K^{(0)}(x_3;\eta_2)\right\},
\end{align*}
where the kernels $K^{(0)}$ and $K^{(k)}$, $k\geq 1$ are defined in \eqref{Qcond3} and \eqref{estK2} respectively. As in the case $m=2$, we now use the assumption that $\nu(x) \leq C \ov{\nu}(x)$ with $\ov{\nu}$ as in \eqref{polykernel} for $\alpha >d$ and some constant $C > 0$. Then the kernels $K^{(k)}$, $0\leq k \leq n$ are estimated as in \eqref{Kkineq}. Therefore,
\begin{equation*}
\begin{split}
&\widetilde{Q}_{3,2}(\eta_1;\eta_2;\eta_3 \mid k) \leq l_1 C^{2} (1+C)^{l_2+l_3-2} (h \ov{\nu}_1 2^{1+\alpha} C)^k \\
\times &\left\{l_2 (l_1 + l_2) l_3 \max_{x_1 \in \eta_1;\,x_2 \in \eta_2} \ov{\nu}(x_1-x_2) \max_{x \in \eta_1 \cup \eta_2;\,x_3 \in \eta_3} \ov{\nu}(x-x_3) \right.\\
&+l_3 (l_1 + l_3) l_2 \max_{x_1 \in \eta_1;\,x_3 \in \eta_3} \ov{\nu}(x_1-x_3) \max_{x \in \eta_1 \cup \eta_3;\,x_2 \in \eta_2} \ov{\nu}(x-x_2) \\
&\left.+ l_2^2 l_3  \max_{x_1 \in \eta_1;\,x_2 \in \eta_2} \ov{\nu}(x_1-x_2)
\max_{x'_2 \in\eta_2;\,x_3 \in \eta_3} \ov{\nu}(x'_2-x_3) +
l_3^2 l_2 \max_{x_1 \in \eta_1;\,x_3 \in \eta_3} \ov{\nu}(x_1-x_3)
\max_{x'_3 \in\eta_3;\,x_2 \in \eta_2} \ov{\nu}(x_2-x'_3)\right\}.
\end{split}
\end{equation*}
It remains to sum over the external trees, see \eqref{treesum}, and we obtain,
\begin{multline}
\label{rho32}
\sum_{n=1}^\infty \frac{1}{n!} Q_{3,2}(\eta_1;\eta_2;\eta_3 \mid n) \\
\leq 2 l_1 l_2 l_3 (l_{1}+l_{2}+l_{3}) C^2 (1+C)^{l_2+l_3-2} \left(\frac{h}{1-h \nu_1 \mathrm{e}}\right)^l
\frac{h \ov{\nu}_1 2^{1+\alpha} C}{1 - h \nu_1 \mathrm{e} - h \ov{\nu}_1 2^{1+\alpha}  C}
\max_{T_{3} \in \mathcal{T}_{3}} \overline{\nu}_{T_{3}}.
\end{multline}
Here, we assumed that $ h (\nu_1 \mathrm{e} + \ov{\nu}_1 2^{1+\alpha} C) < 1$. \\
\indent There remains the case where there is no edge between any points of $\eta_1 \cup \eta_2 \cup \eta_3$.
As explained earlier,
\begin{equation*}
\widetilde{Q}_{3,3}(\eta_1;\eta_2;\eta_3 \mid k) = \sum_{x_1 \in \eta_1}
\sum_{\substack{k_1,k_2,k_3 \geq 0 \\ k_1 + k_2 + k_3 = k-1}} h \int_{\mathbb{R}^{d}} K^{(k_1)}(x_1,y) K^{(k_2)}(y;\eta_2) K^{(k_3)}(y;\eta_3)\, dy,\quad 0 \leq k \leq n,
\end{equation*}
where we set $K^{(k)}(x,y) := K^{(k)}(y;\{x\})$. Inserting the bound \eqref{Kkineq}, we get,
\begin{multline*}
\widetilde{Q}_{3,3}(\eta_1;\eta_2;\eta_3 \mid k) \\
\begin{split}
\leq &C^3 (1+C)^{l_2 +l_3 -2} \sum_{x_1 \in \eta_1} \sum_{\substack{k_1,k_2,k_3 \geq 0 \\ k_1 + k_2 + k_3 = k-1}} (h \ov{\nu}_1 2^{1+\alpha} C)^{k-1}  h \int_{\mathbb{R}^{d}}
\ov{\nu}(x_1-y) \prod_{i=2}^{3} \sum_{x_i \in \eta_i} \ov{\nu}(x_i-y) \, dy \\
\leq &l_1 l_2 l_3 2^{2 \alpha} C^3 (1+C)^{l_2 +l_3 -2}  h \ov{\nu}_1 \sum_{\substack{k_1,k_2,k_3 \geq 0 \\ k_1 + k_2 + k_3 = k-1}} (h \ov{\nu}_1 2^{1+\alpha}  C)^{k-1} \\
&\times \max_{x_1 \in \eta_1;\,x_2 \in \eta_2;\,x_3 \in \eta_3} \left\{ \ov{\nu}(x_1-x_2) \ov{\nu}(x_1-x_3) + \ov{\nu}(x_1-x_2) \ov{\nu}(x_2-x_3) + \ov{\nu}(x_1-x_3) \ov{\nu}(x_2-x_3) \right\}.
\end{split}
\end{multline*}
Summing over trees attached to points of these paths, summing over $n':=n-k$ and using \eqref{treesum},
\begin{align}
&\sum_{n=1}^\infty \frac{1}{n!} {Q}_{3,3}(\eta_1;\eta_2;\eta_3 \mid n) \nonumber \\
&\leq  3 l_1 l_2 l_3 2^{2\alpha} C^3 (1+C)^{l_2 +l_3-2} \sum_{n=1}^\infty \sum_{k=1}^n \frac{h^l (h \ov{\nu}_1)}{(n-k)!} N_{n-k}^{(1)}(k+l) (h \nu_1)^{n-k} \sum_{\substack{k_1,k_2,k_3 \geq 0 \\ \sum\limits_{i=1}^{3} k_i = k-1}} (h \ov{\nu}_1 2^{1+\alpha}C)^{k-1}\max_{T_{3} \in \mathcal{T}_{3}} \overline{\nu}_{T_{3}} \nonumber\\
&\leq 3 l_1 l_2 l_3 2^{2\alpha} C^3 (1+C)^{l_2+l_3-2}
\frac{h^l (h \ov{\nu}_1)}{(1-h \nu_1 \mathrm{e})^{l+1}} \sum_{k=1}^\infty \sum_{\substack{k_1,k_2,k_3 \geq 0 \\ \sum\limits_{i=1}^{3} k_i = k-1}} \left(\frac{h \ov{\nu}_1  2^{1+\alpha}  C}{1-h \nu_1 \mathrm{e}}\right)^{k-1} \max_{T_{3} \in \mathcal{T}_{3}} \overline{\nu}_{T_{3}} \nonumber \\
\label{rho33}
&\leq 3 l_1 l_2 l_3 2^{2 \alpha} C^3 (1+C)^{l_2+l_3-2} \left(\frac{h}{1- h \nu_1 \mathrm{e}}\right)^l \frac{h \ov{\nu}_1 (1-h\nu_1 \mathrm{e})^2}{(1-h\nu_1 \mathrm{e} -  h \ov{\nu}_1 2^{1+\alpha} C)^3} \max_{T_{3} \in \mathcal{T}_{3}} \overline{\nu}_{T_{3}}.
\end{align}
In view of \eqref{rho31}, \eqref{rho32} and \eqref{rho33}, Theorem \ref{polynom} in the case $m=3$ is proven by setting
\begin{gather}
\label{A31}
A_{3,1} := l_1 l_2 l_3 l  C^2  (1+C)^{l_2+l_3-2}\left(\frac{h}{1-h \nu_1 \mathrm{e}}\right)^l, \\
\label{A32}
A_{3,2} := 2 l_1 l_2 l_3 l C^2 (1+C)^{l_2+l_3-2} \left(\frac{h}{1-h \nu_1 \mathrm{e}}\right)^l
\frac{h \ov{\nu}_1 2^{1+\alpha} C}{1- h \nu_1 \mathrm{e} - h \ov{\nu}_1 2^{1+\alpha} C}, \\
\label{A33}
A_{3,3} := 3 l_1 l_2 l_3  2^{2 \alpha} C^3 (1+C)^{l_2+l_3-2} \left(\frac{h}{1-h \nu_1 \mathrm{e}}\right)^l
\frac{h \ov{\nu}_1 (1- h \nu_1 \mathrm{e})^2}{(1- h \nu_1 \mathrm{e} - h \ov{\nu}_1 2^{1+\alpha} C)^3}.
\end{gather}

\subsection{The case of general $m$.}
\label{m=gen}

As before, we integrate out intermediate points $y$, which connect to only  two others (as well as trees of points $y$ connected to a single point of $\bigcup_{i=1}^{m} \eta_i$). We are then left with trees where each $y$ has order $\geq 3$. In terms of Proposition \ref{estQ_m}, we have,
\begin{equation}
\label{generalQm}
Q_m(\eta_1;\dots;\eta_m\mid n) = h^l \sum_{k=0}^n {n \choose k} k! N^{(1)}_{n-k}(l+k) (h\nu_1)^{n-k} \widetilde{Q}_m(\eta_1;\dots;\eta_m \mid k),\quad n \in \mathbb{N}_{0},
\end{equation}
with $\widetilde{Q}_m(\eta_1;\dots;\eta_m \mid k)$ the contributions from trees with $k$ vertices in $\ga$, each of which has order $\geq 2$.\\
\indent Denote by $\sigma$ the number of connected components in $\bigcup_{i=1}^{m}\eta_i$. If $\sigma=1$, i.e. all the $\eta_i$ are connected directly, then there is no such tree, and the contribution is only from $k=0$
\begin{equation*}
\widetilde{Q}_{m,1}(\eta_1;\dots;\eta_m \mid 0) = \sum_{\w{f} \in \mathfrak{S}(\eta_1;\dots;\eta_m\mid \emptyset)} \prod_{(i,j) \in \w{f}} \nu(x_i-x_j).
\end{equation*}
This can be bounded by $C^{l-l_1} N_0^{(m)}(l_1;\dots;l_m) \times \max_{T_{m} \in \mathcal{T}_{m}}\overline{\nu}_{T_{m}}$, but a better bound is obtained as in the case $m=3$, replacing the factor $L_i$ in $N_0^{(m)}(l_1;\dots;l_m)$ by the sum in \eqref{fansum0}
\begin{equation*}
\widetilde{Q}_{m,1}(\eta_1;\dots;\eta_m \mid 0) \leq l^{m-2} (\prod_{i=1}^m l_i)  C^{m-1}  (1+C)^{l-l_1-m+1} \max_{T_{m} \in {\cal T}_m} \ov{\nu}_{T_{m}}.
\end{equation*}
By using the estimate \eqref{treesum}, we obtain,
\begin{gather}
\sum_{n=0}^\infty \frac{1}{n!} Q_{m,1}(\eta_1;\dots,\eta_m \mid n) \leq A_{m,1} \max_{T_{m} \in {\cal T}_m} \ov{\nu}_{T_{m}}, \nonumber \\
\label{Am1}
A_{m,1} := l^{m-2} (\prod_{i=1}^m l_i) C^{m-1} (1+C)^{l-l_1-m+1} \left(\frac{h}{1-h \nu_1 \mathrm{e}}\right)^l,
\end{gather}
which agrees with \eqref{rho31} in case $m=3$. If $\sigma=2$, the only possible such tree is a chain connecting one component to the other. In this case, we have $k\geq 1$ and,
\begin{eqnarray*}
\lefteqn{\w{Q}_{m,2} (\eta_1;\dots;\eta_m \mid k)} \non \\  &=&
\sum_{\substack{I_1 \subset \{1,\dots,m\}\\ 1 \in I_1,\, |I_1| < m}} \sum_{x_1 \in \eta_{I_1}}
\sum_{\substack{i_2 \in I_2\\ I_{2}=I_1^c}}  K^{(k)}(x_1;\eta_{i_2}) \non \\ && \qquad \times \sum_{f_1 \in \mathfrak{S}(\eta_1; \eta_{I_1\setminus \{1\}} \mid \emptyset)} \prod_{(x,x') \in f_1} \nu(x-x') \sum_{f_2 \in \mathfrak{S}(\eta_{i_2}; \eta_{I_2\setminus \{i_2\}}\mid \emptyset)} \prod_{(x,x') \in f_2} \nu(x-x') \non \\
&\leq &C (h \ov{\nu}_1 2^{1+\alpha} C)^k  \sum_{\substack{\{I_1, I_2\} \in \Pi_{2}(\{1,\dots,m\}) \\1 \in I_1}} \sum_{i_2 \in I_2} (1+C)^{l_{i_2}-1}  \sum_{x_1 \in \eta_{I_1}} \sum_{x_2 \in \eta_{i_2}} \max_{x_2 \in \eta_{i_2}} \ov{\nu}(x_1-x_2) \non \\ && \quad \times \sum_{f_1 \in \mathfrak{S}(\eta_1; \eta_{I_1\setminus \{1\}} \mid \emptyset)} \prod_{(x,x') \in f_1} \nu(x-x') \sum_{f_2 \in \mathfrak{S}(\eta_{i_2}; \eta_{I_2\setminus \{i_2\}} \mid \emptyset)} \prod_{(x,x') \in f_2} \nu(x-x').
\end{eqnarray*}
Here, $\Pi_{2}(\{1,\dots,m\})$ denotes the set of all partitions of $\{1,\dots,m\}$ into 2 non-empty subsets. The latter sums over forests can be estimated, replacing again $L_i$ by the sum in \eqref{fansum0}, as
\begin{equation}
\label{forestsum}
\sum_{f_j \in \mathfrak{S}(\eta_{i_{j}}; \eta_{I_{j}\setminus \{i_{j}\}} \mid \emptyset)} \prod_{(x,x') \in f_j} \nu(x-x') \leq l_{I_{j}}^{|I_j|-2} (\prod_{i \in I_j} l_i) C^{|I_j|-1} (1+C)^{\sum\limits_{i \in I_j \setminus \{i_{j}\}} (l_i-1)} \max_{T_{j} \in {\cal T}(I_j)} \ov{\nu}_{T_{j}}.
\end{equation}
Together with the link $(x_1,x_2)$, we obtain a tree on $\{1,\dots,m\}$ (below we denote $l_{I} := \sum_{i\in I} l_{i}$)
\begin{eqnarray*}
\w{Q}_{m,2} (\eta_1;\dots;\eta_m \mid k) \leq (\prod_{i=1}^m l_i) \sum_{\substack{\{I_1,I_{2}\} \in \Pi_{2}(\{1,\dots,m\}) \\ 1 \in I_1}} \prod_{i=1}^{2} l_{I_i}^{|I_i|-1} C^{m-1} (1+C)^{l-l_1-m+1} (h \ov{\nu}_1 2^{1+\alpha} C)^k   \max_{T_{m} \in \mathcal{T}_m} \ov{\nu}_{T_{m}}.
\end{eqnarray*}
The sum over partitions can in fact be evaluated (see \cite{DRS0}) and yields,
\begin{equation*}
\sum_{\substack{\{I_1,I_{2}\} \in \Pi_{2}(\{1,\dots,m\}) \\ 1 \in I_1}}  \prod_{i=1}^{2} l_{I_i}^{|I_i|-1} = (m-1) l^{m-2},\quad m \geq 2.
\end{equation*}
It remains to sum over $n$ and $k$. Hence,
\begin{gather}
\sum_{n=1}^\infty \frac{1}{n!} Q_{m,2}(\eta_1;\dots;\eta_m \mid n)  \leq A_{m,2} \max_{T_{m} \in {\cal T}_m} \ov{\nu}_{T_{m}},\nonumber \\
\label{Am2}
A_{m,2} := (m-1)(\prod_{i=1}^m l_i) l^{m-2} C^{m-1} (1+C)^{l-l_1-m+1} \left(\frac{h}{1-h \nu_1 \mathrm{e}}\right)^l \frac{h \ov{\nu}_1 2^{1+\alpha} C}{1- h \nu_1 \mathrm{e} - h \ov{\nu}_1 2^{1+\alpha} C}.
\end{gather}
Here, we assumed that $h(\nu_1 \mathrm{e} + \ov{\nu}_1 2^{1+\alpha}C)< 1$. Note that this agrees with \eqref{rho32} in case $m=3$. \\
\indent For $\sigma \geq 3$, there is at least one point $y \in \ga$ which is connected to more than two other vertices. The tree on $\ga$ connecting the different components can again be  reduced to a tree $T$ where all vertices have order $\geq 3$ by integrating out the vertices $y$ of order 2, yielding factors $K^{(k)}$. The number of such vertices in $\ga$ is at most $\sigma-2$, where $\sigma$ is the number of connected components of $\bigcup_{i=1}^{m} \eta_i$. The reduction formula reads
\begin{equation}
\label{Qmc}
\w{Q}_{m,\sigma}(\eta_1;\dots;\eta_m \mid k) = \sum_{r=1}^{\min\{k,\sigma-2\}}  \sum_{T \in \mathcal{T}_{r}} \sum_{\substack{\{I_{j}\}_{j=1}^{\sigma} \in \Pi_{\sigma}(\{1,\dots,m\})\\ 1 \in I_{1}}} \sum_{\pi \in \mathcal{M}^{(3)}(T,\sigma,r)} \w{Q}_{\{I_j\}_{j=1}^\sigma,T,\pi}(r),\quad k \geq 1,
\end{equation}
with
\begin{equation*}
\begin{split}
\w{Q}_{\{I_j\}_{j=1}^\sigma,T,\pi}(r) :=& \frac{h^r}{k!} {k \choose r} (k-r)!  \int_{\mathbb{R}^{dr}} dy_1 \cdots dy_r
\sum_{\substack{(k_{y,y'})_{(y,y') \in T} \\ k_{y,y'} \geq 0,\, \sum\limits_{(y,y') \in T} k_{y,y'} \leq k-r}}  \prod_{(y,y') \in T} K^{(k_{y,y'})}(y,y') \\
&\times \sum_{\substack{k_{1},\dots,k_{\sigma} \geq 0 \\ \sum\limits_{j=1}^\sigma k_j + \sum\limits_{(y,y') \in T} k_{y,y'} = k-r}} \sum_{x_1 \in \eta_1} K^{(k_1)}(x_1,y_{\pi(1)})
\times \sum_{f_1 \in \mathfrak{S}(\eta_{1};\eta_{I_1\setminus\{1\}} \mid \emptyset)} \prod_{(x,x') \in f_1} \nu(x-x') \\
& \times \prod_{j=2}^\sigma \left( \sum_{i_j \in I_j}  K^{(k_j)}(y_{\pi(j)}; \eta_{i_j}) \sum_{f_j \in \mathfrak{S}(\eta_{i_j}; \eta_{I_j \setminus \{i_j\}} \mid \emptyset)}  \prod_{(x,x') \in f_j} \nu(x-x')\right).\nonumber
\end{split}
\end{equation*}
In \eqref{Qmc}, $\mathcal{T}_{r}$ denotes the set of tree graphs on $r$ points, $I_j$ the set of $i$ such that $\eta_i$ belongs to the $j$-th component, $\Pi_{\sigma}(\{1,\dots,m\})$ the set of all partitions $\{I_{j}\}_{j=1}^{\sigma}$ of $\{1,\dots,m\}$ into $\sigma$ non-empty subsets and $\mathcal{M}^{(3)}(T,\sigma,r)$ the set of maps $\pi: \{1,\dots,\sigma\} \to \{1,\dots,r\}$ such that $\vert \{y \in T:(y_i,y) \in T\}\vert + \vert \pi^{-1}(i)\vert  \geq 3$, $i=1,\dots,r$, i.e., each point $y_i$ has at least 3 edges attached in the resulting graph. $\pi \in {\cal M}^{(3)}(T,\sigma,r)$ determines the points of attachment of each component to the tree $T$. We point out that the factor $1/k!$ compensates for $k!$ in \eqref{generalQm}, and the factors $k \choose r$ and $(k-r)!$ then count the number of ways of choosing which $y_i$ are associated with the vertices of $T$ and the number of ways of distributing the remaining $y_i$ over the vertices of order 2. If $q$ is the number of vertices $y \in \ga$ of $T$ connected to at least 3 other points of $\ga$, then the tree $T$ determines $q-1$ edges between these vertices. In addition, there are  $q_e \geq 3q-2(q-1)=q+2$ endpoints. Each intermediate point of the tree must be connected to at least one components of $\bigcup_{i=1}^{m} \eta_i$, whereas each endpoint must be connected to at least two. Let $t$ be the number of intermediate points. Then $\sigma \geq t + 2 q_e$ and, as result, $r = q_e + q + t \leq 2q_e - 2 + t \leq \sigma-2$. It follows that \eqref{Qmc} can be rewritten as
\begin{equation}
\label{wQsubTpi}
\w{Q}_{m,\sigma}(\eta_1; \dots; \eta_m \mid k) = \sum_{r=1}^{\min\{k,\sigma-2\}} \sum_{T \in \mathcal{T}_{r}} \sum_{\substack{\{I_{j}\}_{j=1}^{\sigma} \in \Pi_{\sigma}(\{1,\dots,m\})\\ 1 \in I_{1}}} \sum_{\pi \in \mathcal{M}^{(3)}(T,\sigma,r)} \w{Q}_{\{I_j\}_{j=1}^\sigma,T,\pi}(r).
\end{equation}
The contribution of a given tree $T \in \mathcal{T}_{r}$ with $r$ vertices and an assignment $\pi$ is bounded above by
\begin{multline*}
\w{Q}_{\{I_j\}_{j=1}^\sigma,T,\pi}(r) \leq  \frac{h^r}{r!} C^{\sigma+r-1} (h \ov{\nu}_1 2^{1+\alpha} C)^{k-r}\sum_{\substack{(k_{y,y'})_{(y,y') \in T} \\ k_{y,y'} \geq 0,\, \sum\limits_{(y,y') \in T} k_{y,y'} \leq k-r}}
\sum_{\substack{k_{1},\dots,k_{\sigma} \geq 0 \\ \sum\limits_{j=1}^\sigma k_j + \sum\limits_{(y,y') \in T} k_{y,y'} = k-r}} \\ \times \int_{\mathbb{R}^{dr}} dy_1 \cdots dy_r\, \prod_{(y,y') \in T} \ov{\nu}(y-y') \sum_{x_1 \in \eta_{I_1}} \ov{\nu}(x_1-y_{\pi(1)}) \sum_{f_1 \in \mathfrak{S}(\eta_{1};\eta_{I_1 \setminus\{1\}} \mid \emptyset)} \prod_{(x,x') \in f_1} \nu(x-x') \\ \times \prod_{j=2}^\sigma \left(\sum_{i_j \in I_j}  (1+C)^{l_{i_j}-1} \sum_{x_j \in \eta_{i_j}} \ov{\nu}(x_j-y_{\pi(j)})\sum_{f_j \in \mathfrak{S}(\eta_{i_j};  \eta_{I_j \setminus \{i_j\}} \mid \emptyset)}  \prod_{(x,x') \in f_j} \nu(x-x')\right).
\end{multline*}
Inserting the bound \eqref{forestsum} which holds for all $j \in \{1,\dots,\sigma\}$, we have,
\begin{multline}
\label{wqbound2}
\w{Q}_{\{I_j\}_{j=1}^\sigma,T,\pi}(r) \leq  \frac{h^r}{r!} C^{\sigma+r-1} (h \ov{\nu}_1 2^{1+\alpha} C)^{k-r}
 \sum_{\substack{(k_{y,y'})_{(y,y') \in T} \\ k_{y,y'} \geq 0,\, \sum\limits_{(y,y') \in T} k_{y,y'} \leq k-r}}
\sum_{\substack{k_{1},\dots,k_{\sigma} \geq 0 \\ \sum\limits_{j=1}^\sigma k_j + \sum\limits_{(y,y') \in T} k_{y,y'} = k-r}} \\
\times \int_{\mathbb{R}^{dr}} dy_1 \cdots dy_r\, \prod_{(y,y') \in T} \ov{\nu}(y-y') \left(\sum_{x_1 \in \eta_{I_1}} \ov{\nu}(x_1-y_{\pi(1)}) l_{I_1}^{\vert I_1 \vert -2} C^{\vert I_1 \vert-1} (1+C)^{\sum\limits_{i\in I_1\setminus\{1\}} (l_i-1)} \max_{T_1 \in {\cal T}(I_1)} \ov{\nu}_{T_1}\right) \\
\times \prod_{j=2}^{\sigma} \left(\sum_{x_j \in \eta_{I_j}} \ov{\nu}(x_j-y_{\pi(j)}) l_{I_j}^{\vert I_j \vert -2} C^{\vert I_j \vert-1} (1+C)^{\sum\limits_{i\in I_j} (l_i-1)} \max_{T_j \in {\cal T}(I_j)} \ov{\nu}_{T_j}\right).
\end{multline}
Subsequently, we need to bound the integrals
\begin{equation*}
\int_{\mathbb{R}^{dr}} \prod_{(y,y') \in T} \ov{\nu}(y-y') \prod_{j=1}^{\sigma} \ov{\nu}(x_j-y_{\pi(j)})\, dy_1 \cdots dy_r,
\end{equation*}
where $x_j \in \eta_{I_j}$ and $(y,y')$ is a edge in $T$ between $y_i$ and $y_j$ for some $i,j=1,\dots,r$. Note that it is allowed for $y_{\pi(j)}$ to be equal to $y_{\pi(j')}$ with $j\neq j'$. However, the number of unequal $y_{\pi(j)}$ must be at least twice the number of endpoints of the graph $T$ on $\ga$. To estimate this integral, we integrate over the endpoints of $T$ except the endpoint $\pi(1)$ connected to $I_1$. 
Integrating over an endpoint $y$, Lemma \ref{decaylemma} yields
\begin{equation*}
\int_{\mathbb{R}^{d}} \prod_{i=1}^p \ov{\nu}(x_i-y) \ov{\nu}(y-y')\, dy \leq  2^{\alpha p} \ov{\nu}_1
\left(\sum_{i=1}^p \prod_{\substack{k=1 \\ k \neq i}}^{p} \ov{\nu}(x_i-x_k) \ov{\nu}(x_i - y') + \prod_{i=1}^p \ov{\nu}(x_i-y')\right).
\end{equation*}
Therefore, setting $p=|\pi^{-1}(i)|$, we have,
\begin{multline*}
\int_{\mathbb{R}^{d}} \ov{\nu}(y_i-y'_i) \prod_{j \in \pi^{-1}(i)}  \ov{\nu}(x_j-y_i)\, dy_{i}  \\
\leq  2^{\alpha |\pi^{-1}(i)|} {\ov{\nu}_1} \left(\sum_{j \in \pi^{-1}(i)} \prod_{\substack{
j' \in \pi^{-1}(i) \\ j'\neq j}} \ov{\nu}(x_j-x_{j'}) \ov{\nu}(x_j - y'_i) + \prod_{j \in \pi^{-1}(i)} \ov{\nu}(x_j-y'_i)\right).
\end{multline*}
Inserting this into \eqref{wqbound2}, the first term in brackets combines trees $T_j$ on clusters connected to
the same endpoint $y_i$, i.e. $\pi(j) = i$, into a single tree $T'_i$ connected to $y'_i$. The second term connects all $x_j$ with $\pi(j)=i$ to $y'_i$. Note also that the factors $C^{|I_j|-1}$ combine to give $\prod_{j=1}^\sigma C^{|I_j|-1} = C^{m-\sigma}$ and similarly,
\begin{equation*}
\prod_{j=1}^\sigma (1+C)^{\sum\limits_{i \in I_j\setminus \{1\}} (l_i-1)} = (1+ C)^{l-l_1-m+1}.
\end{equation*}
Thus, we obtain,
\begin{multline}
\label{wQbar}
\w{Q}_{\{I_j\}_{j=1}^\sigma,T,\pi}(r) \leq \frac{h^r}{r!} (\prod_{j=1}^\sigma l_{I_j}^{|I_j|-2} ) C^{m+r-1} (1+C)^{l-l_1-m+1} (h \ov{\nu}_1 2^{1+\alpha} C)^{k-r} \\
\times \big\vert\{ (k_i)_{i=1}^{\sigma+r-1}:\sum_{i=1}^{\sigma+r-1} k_i = k-r \} \big\vert\, \ov{Q}_{\{I_j\}_{j=1}^\sigma,T,\pi}(r),
\end{multline}
where it is understood that $k_{i} \geq 0$ for $i=1,\dots,\sigma + r-1$, and where
\begin{equation}
\label{wwQ}
\ov{Q}_{\{I_j\}_{j=1}^\sigma,T,\pi}(r) :=
\int_{\mathbb{R}^{dr}}  \prod_{(y,y') \in T} \ov{\nu}(y-y') \left(\prod_{j=1}^\sigma \sum_{x_j \in \eta_{I_j}} \ov{\nu}(x_j-y_{\pi(j)}) \max_{T_j \in {\cal T}(I_j)} \ov{\nu}_{T_j}\right)\, dy_1 \cdots dy_r.
\end{equation}
Singling out the endpoints $\partial T$ of $T$ other than $\pi(1)$ and denoting $\Lambda(\pi,T) := \pi^{-1}(\partial T \setminus \{\pi(1)\})$, \eqref{wwQ} can be rewritten as
\begin{multline*}
\begin{split}
\ov{Q}_{\{I_j\}_{j=1}^\sigma,T,\pi}(r) =&  \int_{\mathbb{R}^{dr}} dy_1 \cdots dy_r \prod_{(y,y') \in ((T \setminus \partial T) \cup \{\pi(1)\})} \ov{\nu}(y-y') \sum_{x_1 \in \eta_{I_1}} \ov{\nu}(x_1 - y_{\pi(1)})
\max_{T_1 \in {\cal T}(I_1)} \ov{\nu}_{T_1} \\ &\times \prod_{\substack{j>1 \\ \pi(j) \in (T\setminus \partial T)}}\left(\sum_{x_j \in \eta_{I_j}}  \ov{\nu}(x_j-y_{\pi(j)}) \max_{T_j \in {\cal T}(I_j)} \ov{\nu}_{T_j}\right) \\
&\times \sum_{\substack{\{x_j\}_{j \in \Lambda(\pi,T)} \\ x_j \in \eta_{I_j}}} \prod_{j \in \Lambda(\pi,T)}  \ov{\nu}(x_j-y_{\pi(j)}) \ov{\nu}(y_{\pi(j)}- y'_{\pi(j)}) \max_{T_j \in {\cal T}(I_j)} \ov{\nu}_{T_j}.
\end{split}
\end{multline*}
Integrating out the endpoints $y_{\pi(j)}$, we have,
\begin{multline}
\label{wwQbound}
\ov{Q}_{\{I_j\}_{j=1}^\sigma,T,\pi}(r) \leq \int_{\mathbb{R}^{d \vert T \setminus \partial T \vert}} \prod_{i \in (T\setminus \partial T)} dy_i \prod_{(y,y') \in ((T \setminus \partial T) \cup \{\pi(1)\})} \ov{\nu}(y-y') \sum_{x_1 \in \eta_{I_1}}  \ov{\nu}(x_1-y_{\pi(1)}) \max_{T_1 \in {\cal T}(I_1)} \ov{\nu}_{T_1}
\\ \times \prod_{\substack{j> 1 \\ \pi(j) \in (T\setminus \partial T)}} \left(\sum_{x_j \in \eta_{I_j}}  \ov{\nu}(x_j-y_{\pi(j)})\max_{T_j \in {\cal T}(I_j)} \ov{\nu}_{T_j}\right) \prod_{i \in (\partial T\setminus \{\pi(1)\})} \left\{ 2^{\alpha |\pi^{-1}(i)|} \ov{\nu}_1
\sum_{\substack{\{x_j\}_{j \in \pi^{-1}(i)} \\ x_j \in \eta_{I_j}}} \right. \\ \times  \left. \left(\sum_{j \in \pi^{-1}(i)} \prod_{\substack{j' \in \pi^{-1}(i) \\ j' \neq j}}
\ov{\nu}(x_j-x_{j'}) \ov{\nu}(x_j - y'_i) + \prod_{j \in \pi^{-1}(i)} \ov{\nu}(x_j-y'_i)\right) \max_{T_j \in {\cal T}(I_j)} \ov{\nu}_{T_j}\right\}.
\end{multline}
After this first integration, some of the neighbours $y'_i$ have become endpoints of a reduced tree $T$. We integrate out these points next and proceed this way until $T$ is reduced to a single point. We can write the expression \eqref{wwQbound} in terms of the reduced tree as follows
\begin{equation}
\label{modQov}
\ov{Q}_{\{I_j\}_{j=1}^\sigma,T,\pi}(r) \leq 2^{\alpha \vert \pi^{-1}(\partial T \setminus \{\pi(1)\})\vert} \ov{\nu}_1^{|\partial T \setminus \{\pi(1)\}|}
\sum_{\substack{\{I'_{j}\}_{j=1}^{\sigma'} \in \Pi_{\sigma'}(\{1,\dots,m\}) \\ 1 \in I_{1}'}} \sum_{\pi'} \ov{Q}_{\{I'_{j}\}_{j=1}^{\sigma'}, T\setminus \partial T,\pi'}(r),
\end{equation}
where for each $i' \in ((T\setminus \partial T) \cup \{\pi(1)\})$, the set of $j'$ such that $\pi'(j') = i'$ is given by
\begin{equation*}
(\pi')^{-1}(i') = \pi^{-1}(i') \cup \bigcup_{\substack{i \in \partial T \\ y'_i = y_{i'}}} S_i,
\end{equation*}
where, either $S_i := \{j\}$ for some $j \in \pi^{-1}(i)$, or $S_i := \pi^{-1}(i)$. These two cases respectively correspond to the two terms in the last factor of the right-hand side of \eqref{wwQbound}. In the first case, the trees $T_j$ with $\pi(j)=i$ combine into a single tree
\begin{equation*}
T'_{j} := \bigcup_{j' \in \pi^{-1}(i)} T_{j'} \cup\{(x_j,x_{j'}): j' \in \pi^{-1}(i),\,j' \neq j\}.
\end{equation*}
In the second case, the forests $f_j$ are unchanged. The number of components is reduced to
\begin{equation*}
\sigma' = \sigma - \sum_{\substack{i \in \partial T \\ |S_i|=1}} (|\pi^{-1}(i)|-1).
\end{equation*}
The corresponding subdivision is $I'_{j} = \bigcup_{j' \in \pi^{-1}(i)} I_{j'}$ in the first case, and  $I'_j = I_j$ for all $j \in \pi^{-1}(i)$ in the second case. In any case, we obviously have $I'_{j} = I_{j}$ for $j \in \pi^{-1}(i')$. We stress the point that the definition of $\ov{Q}$ in the right-hand side of \eqref{modQov} has been slightly modified: we replaced $\ov{\nu}_{T_j}$ by the quantity
\begin{equation*}
\widetilde{\nu}_{T'_j} := \prod_{\substack{j' \in \pi^{-1}(i) \\ j'\neq j}} l_{I_{j'}} \ov{\nu}_{T'_j} \quad  \textrm{if $S_i = \{j\}$}.
\end{equation*}
After at most $r-1$ stages, the forest graph reduces to a single point $r=1$. At the final stage, we need to integrate over the last vertex $y=y_{\pi(1)}$. Noticing that $I'_1=I_1$, it follows,
\begin{multline*}
\ov{Q}_{\{I'_j\}_{j=1}^{\sigma'},\{y\},1}(r) = \int_{\mathbb{R}^{d}} \sum_{x_1 \in I'_1} \ov{\nu}(x_1-y) \max_{T_1 \in {\cal T}(I_1)} \ov{\nu}_{T_1} \sum_{\substack{\{x_j\}_{j=2}^{\sigma'} \\ x_j \in \eta_{I'_j}}} \prod_{j=2}^{\sigma'} \ov{\nu}(x_j-y) \max_{T'_j \in {\cal T}(I'_j)} \widetilde{\nu}_{T'_j} \\
\leq 2^{\alpha(\sigma'-1)} \ov{\nu}_1  \sum_{\substack{\{x_j\}_{j=1}^{\sigma'} \\ x_j \in \eta_{I'_j}}} \sum_{j=1}^{\sigma'} \left(\prod_{\substack{j'=1 \\ j'\neq j}}^{\sigma'} \ov{\nu}(x_j-x_{j'}) \max_{T \in {\cal T}(\{1,\dots,m\}\setminus\{j\})} \widetilde{\nu}_T\right) \leq 2^{\alpha (\sigma'-1)} \ov{\nu}_1 (\prod_{j=1}^{\sigma} l_{I_j}) \max_{T_{m}\in {\cal T}_m} \ov{\nu}_{T_{m}}.
\end{multline*}
Since all trees generated are distinct, we can write, bounding $\sigma'$ by $\sigma$ at each stage,
\begin{equation*}
\ov{Q}_{\{I_j\}_{j=1}^{\sigma},T,\pi}(r) \leq 2^{\alpha (\sigma r-1)} \ov{\nu}_1^r (\prod_{j=1}^\sigma l_{I_j}) \max_{T_{m} \in {\cal T}_m} \ov{\nu}_{T_{m}}.
\end{equation*}
Inserting this in \eqref{wQbar} and \eqref{wQsubTpi}, we obtain the upper bound
\begin{multline*}
\w{Q}_{m,\sigma}(\eta_1;\dots; \eta_m \mid k) \leq  \sum_{\substack{\{I_{j}\}_{j=1}^{\sigma} \in \Pi_{\sigma}(\{1,\dots,m\})\\ 1 \in I_{1}}} ( \prod_{j=1}^{\sigma} l_{I_j}^{|I_j|-1})  C^{m-1} (1+C)^{l-l_1-m+1} (h \ov{\nu}_1 C)^k \\
\times \sum_{r=1}^{\min\{k,\sigma-2\}} \sum_{T \in \mathcal{T}_{r}} \sum_{\pi \in \mathcal{M}^{(3)}(T,\sigma,r)} \frac{1}{r!} 2^{\alpha(r\sigma-1)}
2^{(1+\alpha)(k-r)} \big\vert\{(k_i)_{i=1}^{\sigma+r-1}:
\sum_{i=1}^{\sigma+r-1} k_i = k-r \}\big\vert \max_{T_{m} \in {\cal T}_m} \ov{\nu}_{T_{m}},
\end{multline*}
where it is understood that $k_{i} \geq 0$ for $i=1,\dots,\sigma + r-1$. We now bound the number of trees $T \in {\cal T}_r$ by $r^{r-2} \leq r! \mathrm{e}^r$, and the number of maps $\pi$ by $r^\sigma \leq (\sigma-2)^\sigma$. Using the combinatoric identity (see \cite{DRS0} for a proof),
\begin{equation}
\label{combidt}
\sum_{\substack{\{I_{j}\}_{j=1}^{\sigma} \in \Pi_{\sigma}(\{1,\dots,m\})\\ 1 \in I_{1}}} ( \prod_{j=1}^{\sigma} l_{I_j}^{|I_j|-1})  = {m-1 \choose \sigma-1} l^{m-\sigma},\quad 2 \leq \sigma \leq m,
\end{equation}
where, as previously, $\Pi_{\sigma}(\{1,\dots,m\})$ denotes the set of all partitions $\{I_{j}\}_{j=1}^{\sigma}$ of $\{1,\dots,m\}$ into $\sigma$ non-empty subsets, we get
\begin{multline*}
\w{Q}_{m,\sigma}(\eta_1;\dots; \eta_m \mid k) \leq
(\sigma-2)^\sigma {m-1 \choose \sigma-1} l^{m-\sigma} C^{m-1}(1+C)^{l-l_1 - m +1}   (h \ov{\nu}_1C)^k  \\ \times \sum_{r=1}^{\min\{k,\sigma-2\}} \mathrm{e}^r 2^{\alpha(r\sigma-1)}2^{(1+\alpha)(k-r)}  \big\vert \{(k_i)_{i=1}^{\sigma+r-1}:\sum_{i=1}^{\sigma+r-1} k_i = k-r\} \big\vert \max_{T_{m} \in {\cal T}_m} \ov{\nu}_{T_{m}}.
\end{multline*}
Summing over $k$ and $n$, we have, setting $n':=n-k$ and $k':=k-r$,
\begin{multline*}
\sum_{n=1}^\infty \frac{1}{n!} Q_{m,\sigma}(\eta_1; \dots; \eta_m \mid n) \\
\begin{split}
\leq& (\sigma-2)^\sigma  {m-1 \choose \sigma-1} l^{m-\sigma} C^{m-1} (1+C)^{l-l_1-\sigma+1}  h^l  \sum_{k=1}^\infty (h \ov{\nu}_1 C)^k  \sum_{n'=0}^\infty \frac{(h \nu_1)^{n'}}{n'!}N_{n'}^{(1)}(k+l) \\
&\times \sum_{r=1}^{\min\{k,\sigma-2\}} 2^{(1+\alpha)(k-r)}2^{\alpha (r\sigma-1)} \mathrm{e}^r \big\vert \{(k_i)_{i=1}^{\sigma+r-1}: \sum_{i=1}^{\sigma+r-1} k_i = k-r \}\big\vert \max_{T_{m} \in {\cal T}_m}  \ov{\nu}_{T_{m}} \\
\leq& (\sigma-2)^\sigma  {m-1 \choose \sigma-1} l^{m-\sigma} C^{m-1}(1+C)^{l-l_1-\sigma+1} \sum_{r=1}^{\sigma-2}
(\ov{\nu}_1 C)^r  \left(\frac{h}{1-h \nu_1 \mathrm{e}}\right)^{l+r} \\
&\times  2^{\alpha (r\sigma-1)} \mathrm{e}^r \sum_{k'=0}^\infty \left( \frac{h \ov{\nu}_1 2^{1+\alpha} C}{1-h \nu_1 \mathrm{e}}\right)^{k'} \big\vert \{(k_i)_{i=1}^{\sigma+r-1}: \sum_{i=1}^{\sigma+r-1} k_i = k' \}\big\vert \max_{T_{m} \in {\cal T}_m}  \ov{\nu}_{T_{m}} \\
\leq & (\sigma-2)^\sigma  {m-1 \choose \sigma-1} 2^{\alpha (\sigma-1)^2} l^{m-\sigma} C^{m-1} (1+C)^{l-l_1-\sigma+1} \left(\frac{h}{1-h \nu_1 \mathrm{e}}\right)^l \left(\frac{1-h \nu_1 \mathrm{e}}{1- h \nu_1 \mathrm{e} - h \ov{\nu}_1 2^{1+\alpha} C}\right)^{\sigma-1} \\ &\times
\sum_{r=1}^{\infty} \left(\frac{h \ov{\nu}_1 C \mathrm{e}}{1- h \nu_1 \mathrm{e} - h \ov{\nu}_1 2^{1+\alpha} C}\right)^r  \max_{T_{m} \in {\cal T}_m}  \ov{\nu}_{T_{m}}.
\end{split}
\end{multline*}
Here, we assumed that $h[\nu_1 \mathrm{e} + \overline{\nu}_{1} C(\mathrm{e} + 2^{1+\alpha})]< 1$. Note that to derive the third inequality, the sum over $k'$ is rewritten as a multiple sum over $k_{1},\dots,k_{\sigma+r-1}$ which yields a factor $(\frac{1- h\nu_1 \mathrm{e}}{1- h \nu_1 \mathrm{e} - h \ov{\nu}_1 2^{1+\alpha} C})^{\sigma+r-1}$. Theorem \ref{polynom} in the case $m\geq 4$ is proven by setting, for any $3\leq \sigma \leq m$,
\begin{multline}
\label{Amm}
A_{m,\sigma} := (\sigma-2)^\sigma  {m-1 \choose \sigma-1} 2^{\alpha (\sigma-1)^2} l^{m-\sigma} C^m (1+C)^{l-l_1-\sigma+1} \left(\frac{h}{1-h \nu_1 \mathrm{e}}\right)^l \\
\times \left(\frac{1- h \nu_1 \mathrm{e}}{1- h \nu_1 \mathrm{e} - h \ov{\nu}_1 2^{1+\alpha} C}\right)^\sigma \frac{h \ov{\nu}_1 \mathrm{e} (1- h \nu_1 \mathrm{e} - h \ov{\nu}_1 2^{1+\alpha} C)}{1-h \nu_1 \mathrm{e}- h \ov{\nu}_1 (\mathrm{e} + 2^{1+\alpha})C}.
\end{multline}

\setcounter{secnumdepth}{0}
\section{Acknowledgments.}

The authors would like to thank the referee for cautious reading of the manuscript and insightful comments that helped improve the clarity and readiness of the paper. The second author (A.L.R.) gratefully acknowledges the financial support of the Ukrainian Scientific Project "III-12-16 Research of models of mathematical physics describing deterministic and stochastic processes in complex systems of natural science". He also gratefully acknowledges the kind hospitality of the Dublin Institute for Advanced Studies, School of Theoretical Physics, during his visit.

\small{}
\end{document}